\documentclass[pra,twocolumn,showpacs]{revtex4}
\usepackage{bbm}
\usepackage{amsmath,amssymb,amsfonts}
\usepackage{times,txfonts}
\usepackage{graphicx}
\usepackage{psfrag}

\newcommand{\ave}[1]{\langle #1 \rangle}

\newcommand{\sgn}{{\mbox{sgn}}}
\newcommand{\cum}[1]{\langle\langle #1 \rangle\rangle}

\def\kbar{\protect\@kbar}
\def\@kbar{%
\relax \bgroup
\def\@tempa{\hbox{\raise.73\ht0
\hbox to0pt{\kern.25\wd0\vrule width.5\wd0
height.1pt depth.1pt\hss}\box0}}%
\mathchoice{\setbox0\hbox{$\displaystyle k$}\@tempa}%
{\setbox0\hbox{$\textstyle k$}\@tempa}%
{\setbox0\hbox{$\scriptstyle k$}\@tempa}%
{\setbox0\hbox{$\scriptscriptstyle k$}\@tempa}%
\egroup}

\begin{document}

\title{Quantum mechanical cumulant dynamics near stable periodic orbits in phase
space: Application to the classical-like dynamics of quantum accelerator modes}
\author{R. Bach}
\affiliation{Center for Theoretical Physics, Polish Academy of Sciences,
02-668 Warsaw, Poland} 
\author{K. Burnett}
\affiliation{Clarendon Laboratory, Department of Physics,
University of Oxford, Oxford OX1 3PU, United Kingdom} 
\author{M. B. d'Arcy} 
\affiliation{Atomic Physics Division, National Institute of Standards and
Technology, Gaithersburg, Maryland 20899-8423, USA} 
\author{S. A. Gardiner}
\affiliation{JILA, University of Colorado and National Institute of Standards 
and Technology, Boulder, Colorado 80309-0440, USA} 

\date{\today}
                                                                                                                                              
\begin{abstract}
We formulate a general method for the study of semiclassical-like dynamics in 
stable regions of a mixed phase-space, in order to theoretically study the 
dynamics of quantum accelerator modes. In the simplest case, this involves 
determining solutions, which are stable when constrained to remain pure-state 
Gaussian wavepackets, and then propagating them using a cumulant-based 
formalism. Using this methodology, we study the relative longevity, 
under different parameter regimes, of quantum accelerator modes. Within this 
attractively simple formalism, we are able to obtain good qualitative agreement 
with exact wavefunction dynamics.
\end{abstract}
                                                                                                                                              
                                                                                 
\maketitle

\section{Introduction}
Since their initial discovery \cite{Oberthaler1999},
quantum accelerator modes
\cite{Oberthaler1999,Godun2000,dArcy2001,Schlunk2003a,Schlunk2003b,
dArcy2003,Ma2004,Buchleitner2004}
have proved to be a fascinating example of a robust 
quantum resonance effect, and
an exciting development in atom-optical studies 
\cite{Oberthaler1999,Godun2000,dArcy2001,Schlunk2003a,Schlunk2003b,
dArcy2003,Ma2004,Buchleitner2004,Moore1994,Klappauf1998,Bharucha1999,Ammann1998,
Milner2001,Truscott2000} 
of quantum-nonlinear phenomena \cite{Gutzwiller1990}.
The demonstrated coherence of their formation 
\cite{Schlunk2003a} promises important applications in coherent atom optics 
\cite{Godun2000,Truscott2000,Berman1997}.
The experimental configuration in which they have been observed is
closely equivalent to those used by the group of Raizen
\cite{Moore1994,Klappauf1998,Bharucha1999}, and subsequently by others 
\cite{Ammann1998} in
the study of quantum $\delta$-kicked rotor dynamics, in particular in observing
the prediction of dynamical localization \cite{Fishman1993}.
In a configuration consisting of a laser cooled cloud of freely falling cesium 
atoms subjected to periodic $\delta$-like kicks from a vertically oriented 
off-resonant laser standing wave 
\cite{Oberthaler1999,Godun2000,dArcy2001,Schlunk2003a,Schlunk2003b,dArcy2003,
Ma2004,Buchleitner2004},
quantum accelerator modes are characterized experimentally by a momentum
transfer, linear with kick number, to a substantial fraction (up to $\sim20$\%) 
of the initial cloud of atoms. The dynamical 
system, including the explicit presence
of gravity, we term the $\delta$-kicked accelerator \cite{dArcy2001}.
The experimental observation of quantum accelerator modes
lead to the pioneering formulation by Fishman, Guarneri
and Rebuzzini \cite{Fishman2002} of the strongly quantum-mechanical dynamics in
terms of an effective classical map. This is justified by the closeness of the
kicking periodicity to particular resonant times. The resulting limiting
dynamics are termed $\epsilon$-classical \cite{Fishman2002,Wimberger2003}, 
and have been used to great effect in the interpretation and prediction of experimentally
observable quantum accelerator modes \cite{Schlunk2003a,Schlunk2003b,Ma2004,
Buchleitner2004}.

The $\delta$-kicked accelerator is therefore also attractive in that it is 
possible to tune its effective classicality in an accessible regime far from 
the true semiclassical limit, making it an ideal testing ground for 
semiclassical theories. 
Semiclassical approaches in quantum chaotic dynamics have proved very 
successful in forging conceptual links between classically chaotic systems and 
their quantum mechanical counterparts \cite{Gutzwiller1990}. When trying to 
include quantum mechanical effects, an obvious step beyond point-particle 
dynamics is to consider the evolution of Gaussian wavepackets. Straightforward 
semiclassical Gaussian wavepacket dynamics are limited in that, e.g., the 
wavepacket is unrealistically forced to maintain its Gaussian form. Pioneering 
work by Huber, Heller, and Littlejohn \cite{Huber1987} proposed remedying this 
by allowing complex classical trajectories. These also permit the study of a 
wider range of classically forbidden processes, and the propagation of 
superpositions of Gaussians. We propose an alternative approach, which, 
most simply, is to follow the dynamics of the cumulants of initially 
Gaussian wavepackets. When taken to second order, the dynamics are described 
purely in terms of means and variances, as in a Gaussian wavepacket, but 
evolution into non-Gaussian wavepackets is not prevented. 
In this paper we develop an appropriate general formalism, and apply it to 
the phenomenon of quantum accelerator modes, thus reaching beyond an
$\epsilon$-classical description of the dynamics.

The Paper is organized as follows: Section II details in a general way the
necessary essential formalism on Gaussian wavepackets and non-commutative
cumulant hierarchies, as well as specifying important differences between
Gaussian wavepacket dynamics and second-order truncations of the cumulant
hierarchy. Section III introduces quantum accelerator modes, as appearing in an
atom-optical $\delta$-kicked accelerator. This is followed by a comprehensive
derivation of the kick-to-kick operator dynamics, crucial for the
second-order treatments that follow. The section culminates with the derivation
of the relevant $\epsilon$-classical dynamics, 
along the lines of Fishman, Guarneri, and Rebuzzini \cite{Fishman2002}.
Section IV builds on the operator dynamics derived in the previous section to
produce an approximate second-order cumulant description, in which
the $\epsilon$-classical theory is a first-order expansion within
the cumulant hierarchy. There follows a worked example of the necessary 
methodology to determine approximate periodic orbits within the resultant
second-order kick-to-kick map, considering the most significant quantum accelerator
modes \cite{Oberthaler1999}. These approximate solutions are propagated with
the second-order mapping equations and with the exact time-evolution operator,
which yields useful insight into the experimentally observed finite lifetimes of
quantum accelerator modes, demonstrably showing the utility of our second-order
approach. Section V consists of the conclusions, which are
followed by four technical appendices, which serve to make the Paper entirely
self-contained.

\section{Overview of the essential general formalism}

\subsection{Gaussian wavepackets}
\label{Sec:Gaussian}

We consider two conjugate self-adjoint operators: $\hat{\xi}$ and 
$\hat{\zeta}$, such that $[\hat{\xi},\hat{\zeta}]=i\eta$, and a Hamiltonian described in
terms of these operators
$\hat{H}(\hat{\xi},\hat{\zeta})$. The dynamics of these operators can be 
fully described by the expectation values $\mu_{\xi}=\langle 
\hat{\xi} \rangle$ and $\mu_{\zeta}=\langle \hat{\zeta} \rangle$ only as 
$\eta\rightarrow 0$. In this limit there is a well-defined $\xi$, $\zeta$ 
phase space, which generally consists of a mixture of stable islands based 
around stable periodic orbits, and a chaotic sea. This is the case for 
the specific system we consider, the $\delta$-kicked accelerator  (see Fig.\ 
\ref{Fig:Poincare}) \cite{dArcy2001}. 

When considering dynamics near a stable periodic orbit in phase space, we note the facts
that: local dynamics approximate those of a harmonic oscillator 
\cite{Lichtenberg1992}, and Gaussian wavepackets remain Gaussian when 
experiencing harmonic dynamics. This can be used as a motivation for 
the initial use of a Gaussian ansatz of the form 
\cite{Huber1987,CommentGaussian}
\begin{equation}
\begin{split}
\psi(\xi) = & 
(2\pi\sigma_{\xi}^{2})^{-1/4} 
\\&\times
\exp\left(-
\frac{[1-i2\sigma_{\xi\zeta}^{2}/\eta][\xi-\mu_{\xi}]^{2}}{4\sigma_{\xi}^{2}}
+\frac{i\mu_{\zeta}[\xi-\mu_{\xi}]}{\eta}
\right),
\label{Eq:GaussAnsatz}
\end{split}
\end{equation}
where
$\sigma_{\xi}^{2} = 
\langle \hat{\xi}^{2} \rangle -\langle \hat{\xi} \rangle^{2}$ 
is the variance in $\hat{\xi}$, and 
$\sigma_{\xi\zeta}^{2} = 
\langle \hat{\xi}\hat{\zeta}+\hat{\zeta}\hat{\xi} \rangle/2 -\langle \hat{\xi} 
\rangle \langle \hat{\zeta} \rangle$
is the symmetrized covariance in $\hat{\xi}$ and $\hat{\zeta}$. 
As Eq.\ (\ref{Eq:GaussAnsatz}) describes a minimum uncertainty wavepacket, the 
$\hat{\zeta}$ variance, 
$
\sigma_{\zeta}^{2} = 
\langle \hat{\zeta}^{2} \rangle -\langle \hat{\zeta} \rangle^{2}
$,
can be deduced from the general uncertainty relation 
\begin{equation}
\sigma_{\xi}^{2}\sigma_{\zeta}^{2} - (\sigma_{\xi\zeta}^{2})^{2} =
\frac{\eta^{2}}{4}.
\label{Eq:Uncertainty}
\end{equation}
This can be seen from Eq.\ (\ref{Eq:GaussAnsatz}), using 
$-i\eta\partial/\partial\xi$ as the $\xi$ representation of $\hat{\zeta}$.
If the stable islands around the periodic orbits of interest are significant 
compared to the size of a minumum uncertainty wavepacket, we may find stable 
periodic orbits in $\mu_{\xi}$, $\mu_{\zeta}$, $\sigma_{\xi}^{2}$, and
$\sigma_{\xi\zeta}^{2}\}$ when such a Gaussian ansatz is enforced. 
In general this stability can only be approximate, but we will nevertheless 
utilize such solutions, as they are good estimates to maximally stable 
Gaussian wavepackets. We note that it is possible to significantly extend
semiclassical techniques to obtain good error estimates for Guassian evolutions,
including for classically chaotic situations \cite{Combescure1997}. Such
calculations require substantial sophistication, and our intent here is somewhat
different, in that we wish to determine useful information within a very simple
description.

\subsection{Non-commutative cumulants}

A complete picture of the observable dynamics can only be determined from the 
time-evolution of all possible expectation values of products of the dynamical 
variables. Except for very simple systems, this produces a 
complicated hierarchy of coupled equations. 

In order to gain any insight we 
must determine a truncation scheme to reduce this to a managable description. 
This is in a sense 
achieved by the Gaussian ansatz, which considers only means 
and variances. Means and variances are the first two orders of an 
infinite hierarchy of cumulants \cite{Gardiner1996}, which we denote by 
double angle brackets to distinguish them from expectation values. The 
non-commutative cumulants can be obtained directly in terms of operator
expectation values through \cite{Fricke1996}
\begin{equation}
\cum{\hat{q}_{1}\cdots\hat{q}_{n}} =
\left.\frac{\partial}{\partial \tau_{1}}\cdots\frac{\partial}{\partial \tau_{n}}
\ln \langle
e^{\tau_{1}\hat{q}_{1}}\cdots
e^{\tau_{n}\hat{q}_{n}}\rangle\right|_{\tau_{1}=0,\ldots,\tau_{n}=0},
\end{equation}
where $\hat{q}_{i}\in \{\hat{\xi},\hat{\mathcal{\zeta}}\}$. More conveniently, the expectation values can be expressed in terms of cumulants:
\begin{equation}
\begin{split}
\langle \hat{q_{1}} \rangle = & \langle\langle \hat{q}_{1} \rangle\rangle,
\\
\langle  \hat{q}_{1}\hat{q}_{2} \rangle = &
\langle \langle  \hat{q}_{1}\hat{q}_{2} \rangle\rangle +
\langle \langle  \hat{q}_{1}\rangle\rangle
\langle \langle \hat{q}_{2} \rangle\rangle, 
\\
\langle  \hat{q}_{1}\hat{q}_{2}\hat{q}_{3} \rangle= &
\langle \langle  \hat{q}_{1}\hat{q}_{2}\hat{q}_{3} \rangle\rangle +
\langle \langle \hat{q}_{1} \rangle\rangle
\langle \langle  \hat{q}_{2}\hat{q}_{3}\rangle\rangle +
\langle \langle \hat{q}_{2} \rangle\rangle 
\langle \langle  \hat{q}_{1}\hat{q}_{3}\rangle\rangle
\\ &
+\langle \langle \hat{q}_{3} \rangle\rangle 
\langle \langle  \hat{q}_{1}\hat{q}_{2}\rangle\rangle +
\langle \langle  \hat{q}_{1}\rangle\rangle
\langle \langle \hat{q}_{2} \rangle\rangle 
\langle \langle \hat{q}_{3} \rangle\rangle,
\\ \vdots,
\label{eq:correlation}
\end{split}
\end{equation}
where the ordered observables have been partitioned in all possible ways into 
products of cumulants. 

Cumulants tend to become smaller with increasing order, 
unlike expectation values or moments, which always increase. 
Intuitively, higher-order cumulants encode only an 
``extra bit'' of information that lower-order cumulants have not yet provided. 
It is therefore often possible to provide a good description by systematically 
truncating, expressing moments of all orders in terms of cumulants up to some 
finite order \cite{Fricke1996,Marcienkiewicz1939}. 
Truncating at first order is equivalent to considering only mean values, and 
thus reproduces the corresponding Hamilton's equations 
of motion. 

It is tempting to think that truncating at second order is 
equivalent to enforcing a Gaussian ansatz. As will be shown explicitly in our model system, the $\delta$-kicked
accelerator, this does not in general 
reproduce the dynamics given by enforcing a Gaussian ansatz. Gaussian 
wavepacket dynamics are unitary, meaning that the uncertainty relation of Eq.\ (\ref{Eq:Uncertainty}) 
is always exactly observed, and that one need 
consider only two of 
$\sigma_{\xi}^{2}$, $\sigma_{\zeta}^{2}$, and $\sigma_{\xi\zeta}^{2}$. This is only 
true when no terms in the Hamiltonian are of greater than quadratic order in 
$\hat{\xi}$ and $\hat{\zeta}$ \cite{Huber1987}. Furthermore, finding a fixed 
point of $\mu_{\xi}$, $\mu_{\zeta}$, $\sigma_{\xi}^{2}$, and $\sigma_{\xi\zeta}^{2}$ is 
equivalent to finding a perfectly Gaussian eigenstate of the system, 
which can only be true for the harmonic oscillator. 

When propagating the second-order truncated equations of motion for the first 
and second order cumulants, it is  
necessary to consider the dynamics 
of each of $\sigma_{\xi}^{2}$, $\sigma_{\zeta}^{2}$, and $\sigma_{\xi\zeta}^{2}$ 
explicitly, as the uncertainty relation is not explicitly hard-wired into the formalism.
This implies that the evolution described solely in terms of the first and second 
order cumulants is not unitary. This feature of this approach more 
accurately reflects the fact that truncating generally leaves us with an 
incomplete description of the dynamics, with a correspondingly inevitable loss 
of information about the state of the system. We are not, in principle, restricted to 
initially pure states, although this flexibility is not exploited in this paper.

Nonetheless, when situated inside a stable island in $\xi$, $\zeta$ phase 
space, such a ``stable'' Gaussian wavepacket should be long-lived due to the 
harmonic nature of the local dynamics \cite{Tomsovic1991}. We can then use 
the equations of motion appropriate to second-order cumulant dynamics to get 
an idea of how long-lived the initial wavepacket actually is, as physically 
sensible imperfections are included in the dynamics in a straightforward 
manner. 

The approach we have outlined is most obviously applicable in the standard 
semiclassical regime, but is not restricted to it. We will illustrate our 
method by applying it to a very interesting and experimentally relevant system, 
the quantum $\delta$-kicked accelerator \cite{dArcy2001}, 
outside the semiclassical regime (see, also, Appendix \ref{App:Classical}). 
Our approach provides 
useful insights on the longevity of 
quantum accelerator modes in this system, 
essential for their possible application in coherent atom optics 
\cite{Godun2000,Berman1997}. 

\section{Quantum accelerator modes}

\subsection{Introduction to the $\delta$-kicked accelerator}

\subsubsection{Experiment}

In the Oxford experimental realization of the quantum $\delta$-kicked accelerator,
$\sim 10^7$ cesium atoms are trapped and cooled in a MOT (magneto-optical trap) 
to a
temperature of $5\mu$K, yielding a Gaussian momentum distribution
with a full width half maximum of 12 photon recoils \cite{NoteRecoil}. 
The atoms are then released and exposed to a
sequence of equally spaced pulses from a standing wave of higher
intensity light $15$\thinspace GHz red-detuned from the
$6^{2}S_{1/2} \rightarrow 6^{2}P_{1/2}$, ($F=4\rightarrow F'=3$)
D1 transition. 

In this parameter regime, 
the spatial period of the standing wave is
$\lambda_{\mbox{\scriptsize spat}}=447$\thinspace nm, and the half-Talbot time
$T_{1/2}=66.7$\thinspace $\mu$s.
The half-Talbot time is equal to the pulse periodicity at which a quantum antiresonance
would be observed if the initial condition were a zero-momentum plane wave
\cite{dArcy2001,Bharucha1999}, and is so-named in analogy to the Talbot length 
in classical optics \cite{Godun2000,Longhurst1962}. This quantity is of central importance, as it is when the pulse
periodicity approaches integer multiples of $T_{1/2}$ that quantum accelerator modes are
observed experimentally \cite{Oberthaler1999,Godun2000,dArcy2001,Schlunk2003a,
Schlunk2003b,dArcy2003,Ma2004}.
The peak intensity in the standing wave is $\simeq 5 \times 10^4$\thinspace
mW/cm$^2$, and the pulse duration is $t_{p}=500$\thinspace ns.
This is sufficiently short that the atoms are in the Raman-Nath
regime for the observed momentum scales, 
and hence each pulse is a good approximation to a
$\delta$-function kick. The potential depth is quantified by
$\phi_d=\Omega^2 t_p/8\delta_L$, where $\Omega$ is the Rabi
frequency and $\delta_{L}$ the detuning from the D1 transition. 

In the Oxford experiment,
the widths of both the beam and the initial laser-cooled atomic cloud are $\sim 1$\thinspace
mm, and situations exist where the consequent inhomogeneity of the driving strength must be
taken explicitly into account \cite{Schlunk2003a}. It is frequently sufficient to consider
an averaged value only, however \cite{dArcy2001}, which is the approach we adopt in this
Paper. This is also appropriate for a configuration where the initial atomic cloud is
much smaller than the laser beam profile.

After the
pulsing sequence, the atoms fall through a sheet of laser light
resonant with the $6^{2}S_{1/2}\rightarrow 6^{2}P_{3/2}$,
$(F=4\rightarrow F''=5)$ D2 transition, $0.5$\thinspace m below
the MOT. By monitoring the absorption, the atoms' momentum
distribution can then be measured by a time-of-flight method, with a
resolution of 2 photon recoils. For more complete details of the Oxford 
experimental configuration, see Refs.\
\cite{dArcy2001,Godun2000}.

\subsubsection{Model Hamiltonian}

The dynamics of the atoms in the 
Oxford quantum accelerator modes experiment 
\cite{Oberthaler1999,Godun2000,dArcy2001,Schlunk2003a,Schlunk2003b,dArcy2003,
Ma2004} 
are well modelled by the one-dimensional $\delta$-kicked accelerator 
Hamiltonian:
\begin{equation}
\hat{H} = \frac{\hat{p}^2}{2m} + mg\hat{z} - \hbar \phi_d[ 1+\cos(G\hat{z})]
\sum_{n=-\infty}^{\infty} \delta\left( t - n T\right).
\label{Eq:HamiltonPhysical}
\end{equation}
Here $\hat{z}$ is the position, $\hat{p}$ the momentum, 
$m$ the particle mass, $g$ the gravitational acceleration, $t$ the time,
$T$ denotes the pulse period, $G=2\pi/\lambda_{\mbox{\scriptsize spat}}$ where 
$\lambda_{\mbox{\scriptsize spat}}$ is the spatial period of the potential
applied to the atoms, and $\hbar \phi_d$ quantifies the depth of this potential.

As this Hamiltonian is periodic in time, the time evolution of the system can be
described by repeated application of the Floquet operator
\begin{equation}
\hat{F}=\exp
\left(-\frac{i}{\hbar}
\left[
\frac{\hat{p}^{2}}{2m} + mg\hat{z}
\right]T
\right)
\exp\left(
i\phi_{d}
\left[
1+\cos(G\hat{z})
\right]
\right),
\label{Eq:FloquetBase}
\end{equation}
which describes the time evolution from immediately before the application of
one kick to immediately before the application of the next.

\subsubsection{Resonance condition}

The near-fulfilment of the quantum resonance condition (closeness to particular 
resonant pulse periodicities \cite{Godun2000,Fishman2002,Berry1999}) means the free 
evolution of a wavefunction, e.g.\ initially well localized in momentum and 
(periodic) position space immediately after it experiences a kick, causes it 
to rephase to close to its initial condition just before each subsequent kick.

The innovative treatment due to Fishman, Guarneri, and Rebuzzini accounts for
this in terms of a so-called $\epsilon$-classical limit
\cite{Fishman2002,Wimberger2003,Berry1999}, 
where a kind of
kick-to-kick classical point dynamics is regained in the limit of the pulse 
periodicity approaching integer multiples of the half-Talbot time  
$T_{1/2}=2\pi m/\hbar G^{2}$ \cite{Godun2000}, i.e., as $\epsilon =
2\pi(T/T_{1/2}-\ell)\rightarrow 0$, where $\ell \in \mathbb{Z}$. This 
accurately accounts for the
observed acceleration for up to $\sim 100$ kicks, as well as 
predicting numerous experimentally observed high-order 
accelerator modes \cite{Schlunk2003b}. It is this $\epsilon$,
whose smallness indicates 
nearness to special resonant kicking frequencies, leading to the production of 
quantum accelerator modes \cite{Fishman2002}, and
not $\hbar$ (or equivalently the commonly used dimensionless rescaled Planck
constant $\kbar$ \cite{Klappauf1998}, as defined in Appendix
\ref{App:Classical}), which takes the place of $\eta$ in our cumulant-based approach. 

We now describe the treatment of Ref.\ \cite{Fishman2002} to justify the 
appropriate phase-space which is the starting point of our analysis, in a
somewhat different, more operator-oriented form, which turns out to be more
convenient for our purposes.
We provide enough detail for the explanation to be self-contained, as well as setting
notation and a conceptual basis for our second-order cumulant treatment, 
as applied to quantum accelerator mode dynamics.

\subsection{Derivation of the Heisenberg map}

\subsubsection{Gauge transformation}

Moving to a frame comoving with the gravitational
acceleration [$\hat{U}(t)=\exp(img\hat{z}t/\hbar)$], the Hamiltonian of
Eq.\ (\ref{Eq:HamiltonPhysical}) transforms to
$\hat{U}(t)\hat{H}\hat{U}^{\dagger}(t) -mg\hat{z}$. We write the
resulting gauge-transformed Hamiltonian as
\begin{equation}
\hat{H}
=\frac{(\hat{p}-mgt)^{2}}{2m} - \hbar\phi_{d}
[1+\cos(G\hat{z})]\sum_{n=-\infty}^{\infty}\delta(t-nT).
\label{Eq:HamiltonianGauge}
\end{equation}

Due to the explicit time-dependence in the free-evolution part of the
Hamiltonian, it is no longer periodic in time. The Floquet operator of Eq.\
(\ref{Eq:FloquetBase}), which transforms to
$\hat{U}(nT)\hat{F}\hat{U}^{\dagger}([n-1]T)$, consequently also has an explicit
time dependence, and it becomes necessary to specify which kick, and subsequent
free evolution, it is describing. 

We therefore write the gauge-transformed
kick-to-kick time-evolution operator, from $t=(n-1)T$ to $t=nT$, as
\begin{equation}
\begin{split}
\hat{F}_{n}=&\exp\left(
-\frac{i}{\hbar}
\left[
\frac{\hat{p}^{2}}{2m}T-\frac{\hat{p}g}{2}(2n-1)T^{2}
\right]
\right)\\
&\times
\exp\left(
i\phi_{d}
\left[
1+\cos(G\hat{z})
\right]
\right),
\end{split}
\label{Eq:FloquetGauge}
\end{equation}
where we have neglected an irrelevant global phase equal to
$\exp(-i mg^{2}[3n^{2}-3n+1]T^{3}/6\hbar)$. 

The time-evolution operator from
$t=0$ to $t=nT$ is then given by
\begin{equation}
\hat{\mathcal{F}}_{n}=\hat{F}_{n}\hat{F}_{n-1}\ldots\hat{F}_{1} = 
\hat{\mathcal{T}}\left(
\prod_{n'=1}^{n}\hat{F}_{n'}
\right),
\end{equation}
where $\hat{\mathcal{T}}$ is a time-ordering operator.

\subsubsection{Bloch theory}

In this gauge the Hamiltonian is spatially periodic, and we can therefore invoke
Bloch theory \cite{Fishman2002,Ashcroft1976}. Specifically, we parametrize the momentum eigenstates as 
$|(\hbar G)^{-1}p = k + \beta\rangle$, where $k \in \mathbb{Z}$, 
$\beta \in [0,1)$, and $\beta$ is termed the {\it quasimomentum}. The
Hamiltonian's spatial periodicity implies that only momentum eigenstates
differing by integer multiple of $\hbar G$ will couple together, i.e., the
quasimomentum is {\it conserved}.

Correspondingly, the position eigenstates are parametrized as
$|Gz = 2\pi l+\theta\rangle$, where $l \in \mathbb{Z}$, 
$\theta \in [0,2\pi)$. The position and momentum {\it operators\/} can be
similarly parametrized as 
\begin{align}
G\hat{z} = 2\pi\hat{l} + \hat{\theta},
\label{Eq:PositionParam}
\\
(\hbar G)^{-1}\hat{p}  = \hat{k} + \hat{\beta},
\label{Eq:MomentumParam}
\end{align}
defined such that
\begin{align}
\hat{l}|Gz = 2\pi l + \theta\rangle 
& = l|Gz = 2\pi l + \theta\rangle,\\
\hat{\theta}|Gz = 2\pi l + \theta\rangle 
& = \theta|Gz = 2\pi l + \theta\rangle,\\
\hat{k}|(\hbar G)^{-1}p = k + \beta\rangle 
& = k|(\hbar G)^{-1}p = k + \beta\rangle,\\
\hat{\beta}|(\hbar G)^{-1}p = k + \beta\rangle 
& = \beta|(\hbar G)^{-1}p = k + \beta\rangle.
\end{align}

Substituting Eqs.\ (\ref{Eq:PositionParam}) and (\ref{Eq:MomentumParam}) into 
Eq.\ (\ref{Eq:HamiltonianGauge})
the Hamiltonian in the frame accelerating with gravity can therefore be written as
\begin{equation}
\hat{H}=\frac{[\hbar G(\hat{k}+\hat{\beta})-mgt]^{2}}{2m}
-\hbar\phi_{d}[1+\cos(\hat{\theta})]\sum_{n=-\infty}^{\infty}\delta(t-nT),
\end{equation}
where from the commutation relations (see Appendix \ref{App:Bloch} for derivations)
$
[\hat{\beta},\hat{\theta}] = [\hat{\beta},\hat{k}] = 0$, we deduce that
$[\hat{\beta},\hat{H}]=0$, and thus confirm that the quasimomentum is a
conserved quantity {\it in the accelerating frame}.

\subsubsection{Time-evolution operator}

Substituting Eqs.\ (\ref{Eq:PositionParam}) and (\ref{Eq:MomentumParam})
into Eq.\ (\ref{Eq:FloquetGauge}), we get
\begin{equation}
\begin{split}
\hat{F}_{n}=&\exp\left(
-i
\left[
(\hat{k}^{2}+2\hat{k}\hat{\beta})\frac{\hbar
G^{2}T}{2m}-\hat{k}(2n-1)\frac{gGT^{2}}{2}
\right]
\right)
\\&\times
\exp\left(
-i
\left[
\hat{\beta}^{2}\frac{\hbar G^{2}T}{2m}-\hat{\beta}(2n-1)\frac{gGT^{2}}{2}
\right]
\right)
\\&\times
\exp\left(
i\phi_{d}
\left[
1+\cos(\hat{\theta})
\right]
\right),
\end{split}
\end{equation}
where the middle, quasimomentum-dependent term commutes with the the other terms
of the kick-to-kick time-evolution operator, and with all subsequently applied
such operators. After $n$ kicks, the result is a phase determined
entirely by the initial quasimomentum $\beta$, equal to 
$\exp (-in[\beta^{2}\hbar G^{2}T/2m-\beta ngGT^{2}/2])$. 

Removing this
relatively trivial time evolution from explicit consideration, 
recall that we are interested in the case whre $T$ is close to integer multiples of the
half-Talbot time $T_{1/2}=2\pi m/\hbar G^{2}$. Thus, substituting in 
$T=(2\pi\ell+\epsilon)m/\hbar G^{2}$, and using that $e^{-ik^{2}\pi\ell} = 
e^{-ik\pi\ell}$
\begin{equation}
\begin{split}
\hat{F}_{n}=&\exp\left(
-i
\left\{
\frac{\hat{k}^{2}}{2}\epsilon+2\pi\hat{k}
\left[
\frac{\ell}{2}+\hat{\beta}\frac{T}{T_{1/2}}-\left(n-\frac{1}{2}\right)\Omega
\right]
\right\}
\right)
\\&\times
\exp\left(
i\phi_{d}
\left[
1+\cos(\hat{\theta})
\right]
\right),
\end{split}
\label{Eq:FinalFloquet}
\end{equation}
where we have inserted $2\pi \Omega = gGT^{2}$, a dimensionless parameter accounting
for the effect of the gravitational acceleration, also known as the {\it
unperturbed winding number\/} \cite{Buchleitner2004}.

The total time-evolution operator from
$t=0$ to $t=nT$ is then given by
\begin{equation}
\begin{split}
\hat{\mathcal{F}}_{n} &=
 \hat{\mathcal{T}}\left(
\prod_{n'=1}^{n}\hat{F}_{n'}
\right)
\exp 
\left(-i\pi n
\left[
\hat{\beta}^{2}\frac{T}{T_{1/2}}-\hat{\beta} n\Omega
\right]
\right)
\\&=
\exp 
\left(-i\pi n
\left[
\hat{\beta}^{2}\frac{T}{T_{1/2}}-\hat{\beta} n\Omega
\right]
\right)
 \hat{\mathcal{T}}\left(
\prod_{n'=1}^{n}\hat{F}_{n'}
\right).
\end{split}
\end{equation}

\subsubsection{Effective Hamiltonian}

The dynamics governed by the time-evolution operator of 
Eq.\ (\ref{Eq:FinalFloquet}) are, just prior to each kick, exactly equivalent to
those determined by the effective Hamiltonian
\begin{equation}
\begin{split}
\hat{H} =& 
\sum_{n=-\infty}^{\infty}
\Theta(nT-t)\hat{H}_{n}\Theta(t-[n-1]T)
\\&
-\hbar\phi_{d}[1+\cos(\hat{\theta})]
\sum_{n=-\infty}^{\infty}\delta(t-nT)
\end{split}
\label{Eq:EffectiveHamiltonian}
\end{equation}
where $\Theta$ is a step function. Essentially this means that the free
evolution Hamiltonian
\begin{equation}
\hat{H}_{n} =
\frac{\hbar}{T}
\left\{
\frac{\hat{k}^{2}}{2}\epsilon + 2\pi\hat{k}
\left[
\frac{\ell}{2}-
\left(
n-\frac{1}{2}\right)\Omega
+\hat{\beta}\frac{T}{T_{1/2}}
\right]
\right\},
\end{equation}
can be considered time-independent {\it between\/} two successive kicks, but
from before to after application of each kick, there is a stepwise change to
the term linear in $\hat{k}$.

\subsubsection{Heisenberg map}

Using the commutation relations (see Appendix \ref{App:Bloch} for a derivation)
\begin{gather}
[\hat{\theta},\hat{k}]= i, \\
[\hat{k},\hat{l}] = [\hat{l},\hat{\beta}] = \frac{i}{2\pi},\\
[\hat{\theta},\hat{\beta}] =
[\hat{\theta},\hat{l}]= 
[\hat{k},\hat{\beta}]=0,
\end{gather}
the Heisenberg equations of motion for the dynamical variables can be readily 
determined from the effective Hamiltonian given in Eq.\ (\ref{Eq:EffectiveHamiltonian}),
and solved to give a discrete kick-to-kick Heisenberg map:
\begin{subequations}
\begin{align}
\hat{\theta}_{n+1} &=\hat{\theta}_{n} + \epsilon\hat{k}_{n+1} + 2\pi\left[
\frac{\ell}{2}-
\left(
n+\frac{1}{2}
\right)\Omega
+\hat{\beta}_{n}\frac{T}{T_{1/2}}
\right],
\label{Eq:HeisenbergMaptheta}\\
\hat{k}_{n+1} &= \hat{k}_{n} -\phi_{d}\sin(\hat{\theta}_{n}),
\label{Eq:HeisenbergMapk}\\
\hat{l}_{n+1} &= \hat{l}_{n} + \ell\hat{k}_{n+1}-
\left[
\frac{\ell}{2}-
\left(
n+\frac{1}{2}
\right)\Omega
+\hat{\beta}_{n}\frac{T}{T_{1/2}}
\right],
\label{Eq:HeisenbergMapl}\\
\hat{\beta}_{n+1} &=\hat{\beta}_{n}.
\label{Eq:HeisenbergMapbeta}
\end{align}
\label{Eq:HeisenbergMap}
\end{subequations}

Using this procedure to determine Eq.\ (\ref{Eq:HeisenbergMap})
is exactly equivalent to determining the kick-to-kick time-evolution
directly from the time-evolution operator of Eq.\
(\ref{Eq:FinalFloquet}) by $\hat{\theta}_{n+1} =
\hat{F}_{n+1}^{\dagger}\hat{\theta}_{n}\hat{F}_{n+1}$. Equation
(\ref{Eq:HeisenbergMapbeta}) is simply a confirmation that the quasimomentum is
a conserved quantity, and from Eqs.\ (\ref{Eq:HeisenbergMaptheta}) and
(\ref{Eq:HeisenbergMapk}) we see that the evolution of the 
conjugate pair $\hat{\theta}$ and
$\hat{k}$ is completely decoupled from that of the discrete position variable
$\hat{l}$. Both of these properties are a direct consequence of the spatial
periodicity of the gauge-transformed Hamiltonian, combined with the (typical) 
chosen parametrization of $\hat{z}$ and $\hat{p}$ into discrete and continuous
components.  

\subsubsection{Partial trace}
\label{Sec:PartialTrace}

Corresponding to what in Ref.\ \cite{Fishman2002} is termed a {\it Bloch-Wannier
fibration}, it is convenient to exploit the fact that subspaces associated with
distinct values of the quasimomentum $\beta$ are decoupled, by considering the
evolution of the relevant dynamical variables $\hat{\theta}$ and $\hat{k}$, 
conditional to a specific value of the quasimomentum $\beta$, separately.

Starting from a general density operator
\begin{equation}
\begin{split}
\rho =& \iint d\beta d\beta' \sum_{k,k' = -\infty}^{\infty}
d_{kk'}(\beta,\beta')
\\ &\times
|(\hbar G)^{-1}p = k + \beta\rangle
\langle (\hbar G)^{-1}p = k' + \beta'|
\end{split}
\end{equation}
we consider expectation values determined by a form of partial trace, over the
density operator multiplied by the dynamical observable. In general, we define
\begin{widetext}
\begin{equation}
\ave{\hat{\theta}^{a}\hat{k}^{b}\hat{\beta}^{c}}(\beta)
= \frac{1}{\mathcal{N}(\beta)}\sum_{k=-\infty}^{\infty}
\langle (\hbar G)^{-1}p = k + \beta|
\rho \hat{\theta}^{a}\hat{k}^{b}
|(\hbar G)^{-1}p = k + \beta\rangle\beta^{c},
\label{Eq:PTrace}
\end{equation}
\end{widetext}
where 
$\mathcal{N}(\beta)
=\sum_{k=-\infty}^{\infty} d_{kk}(\beta,\beta)
$
is a normalizing constant, causing the expectation
value to be independent of the proportion of the population occupying 
any given $\beta$
subspace.

If we define the projection operator
\begin{equation}
\hat{\mathcal{P}}(\beta) = \sum_{k=-\infty}^{\infty}
|(\hbar G)^{-1}p = k + \beta\rangle\langle (\hbar G)^{-1}p = k + \beta|,
\label{Eq:ProjectionBeta}
\end{equation}
which is that component of the identity specific to a particular
quasimomentum subspace, then we can define the expectation values of interest by
taking the full trace over the density operator multiplied by the operator of
interest and a projector:
\begin{equation}
\ave{\hat{\theta}^{a}\hat{k}^{b}\hat{\beta}^{c}}(\beta)
=\frac{\mbox{Tr}\{\rho
\hat{\theta}^{a}\hat{k}^{b}\hat{\mathcal{P}}(\beta)\}}
{\mbox{Tr}\{\rho \hat{\mathcal{P}}(\beta)\}}\beta^{c},
\label{Eq:Expect}
\end{equation}
where $\mbox{Tr}\{\rho \hat{\mathcal{P}}(\beta)\}\equiv \mathcal{N}(\beta)$.

\subsubsection{Projected Heisenberg map}

Equation (\ref{Eq:Expect}) shows that expectation values, which are in general
time-dependent, of the operator-valued observables
$\hat{\theta}$ and $\hat{k}$, multiplied by the projection operator
$\hat{\mathcal{P}}(\beta)$, are exactly equal to the values determined by taking the
$\beta$-dependent partial traces of $\hat{\theta}$ and $\hat{k}$, considered on their own
(i.e., without a projection operator).

It is therefore clear that the dynamics of $\hat{\theta}$ and 
$\hat{k}$, conditional to a particular value of the quasimomentum $\beta$, can
be equivalently described by the evolution of the projected operators
$\hat{\theta}(\beta)=\hat{\theta}\hat{\mathcal{P}}(\beta)$ and 
$\hat{k}(\beta)=\hat{k}\hat{\mathcal{P}}(\beta)$. 

We can thus speak of a {\it projected Heisenberg map}:
\begin{subequations}
\begin{align}
\begin{split}
\hat{\theta}_{n+1}(\beta) 
=& \hat{\theta}_{n}(\beta) 
+ \mbox{sgn}(\epsilon)\hat{\mathcal{I}}_{n+1}(\beta)\\
&+ 2\pi\left[
\frac{\ell}{2}-
\left(
n+\frac{1}{2}
\right)\Omega
+\beta\frac{T}{T_{1/2}}
\right]\hat{\mathcal{P}}(\beta),
\end{split}
\label{Eq:ProjHeisenbergMaptheta}\\
\hat{\mathcal{I}}_{n+1}(\beta) =& \hat{\mathcal{I}}_{n}(\beta) 
-\tilde{k}\sin(\hat{\theta}_{n})\hat{\mathcal{P}}(\beta),
\label{Eq:ProjHeisenbergMapI}
\end{align}
\label{Eq:ProjHeisenbergMap}
\end{subequations}
where we have introduced $\mathcal{I}=|\epsilon|\hat{k}$ and
$\tilde{k}=|\epsilon|\phi_{d}$. This rescaling implies  the following
commutators: 
\begin{align}
[\hat{\theta},\hat{\mathcal{I}}] =& i|\epsilon|,
\\
[\hat{\theta}(\beta),\hat{\mathcal{I}}(\beta')] =&
i|\epsilon|\delta(\beta-\beta')\hat{\mathcal{P}}(\beta)
\end{align}
which will therefore vanish as $\epsilon\rightarrow 0$, i.e., as $T$ approaches
rational multiples of the half-Talbot time $T_{1/2}$.

If, furthermore, we introduce
\begin{equation}
\hat{\mathcal{J}}_{n}(\beta) = \hat{\mathcal{I}}_{n}(\beta) +\mbox{sgn}(\epsilon)
2\pi\left[
\frac{\ell}{2}-
\left(
n-\frac{1}{2}
\right)\Omega
+\beta\frac{T}{T_{1/2}}
\right]\hat{\mathcal{P}}(\beta),
\label{Eq:Transformation}
\end{equation}
we remove the explicit time-dependence and $\beta$-dependence of the free
evolution, to arrive at
\begin{subequations}
\begin{align}
\hat{\theta}_{n+1}(\beta) =& \hat{\theta}_n(\beta) + 
\sgn(\epsilon)\hat{\cal{J}}_{n+1}(\beta), 
\label{Eq:HeisenbergEq1} \\
\hat{\cal{J}}_{n+1}(\beta) =& \hat{\cal{J}}_{n}(\beta) - \tilde{k} \sin
(\hat{\theta}_n)\hat{\mathcal{P}}(\beta) - 
 \sgn(\epsilon)2\pi\Omega\hat{\mathcal{P}}(\beta).
\label{Eq:HeisenbergEq2}
\end{align}
\label{Eq:Heisenberg}
\end{subequations}
With an appropriate definition of of the inner product, the dynamics can be
mapped onto those for a rotor, the so-called $\beta$-rotors of Ref.\
\cite{Fishman2002} (see Appendix \ref{App:BetaRotor}). 

\subsection{$\epsilon$-classical dynamics}

\subsubsection{Derivation of the $\epsilon$-classical map}

To determine the dynamics of the mean values (conditioned to a particular value
of the quasimomentum $\beta$), one simply takes the appropriately normalized
expectation values of both sides of Eq.\ (\ref{Eq:Heisenberg}). 

Due to the
presence of $\sin(\hat{\theta_{n}})\hat{\mathcal{P}}(\beta)$, this generally
couples to an infinite hierarchy of higher-order moments or cumulants. However,
in the limit $\epsilon\rightarrow 0$, we can make the substitution
\begin{equation}
\frac{\mbox{Tr}
\{\rho\hat{\theta}^{a}{\hat{\mathcal{P}}(\beta)}\}}
{\mbox{Tr}\{\rho\hat{\mathcal{P}}(\beta)\}}
=
\left[\frac{\mbox{Tr}\{\rho\hat{\theta}\hat{\mathcal{P}}(\beta)\}}
{\mbox{Tr}\{\rho\hat{\mathcal{P}}(\beta)\}}
\right]^{a}
=\theta(\beta)^{a}.
\label{Eq:Expectation}
\end{equation}
Essentially the limit $\epsilon\rightarrow 0$ allows us to treat the
$\hat{\theta}(\beta)$, $\hat{\mathcal{J}}(\beta)$ dynamics as those of a point
particle, and higher order correlations can be discarded. This truncation is discussed more
comprehensively in Section \ref{Sec:Second-Order}.

The result is that the dynamics reduce to the kick-to-kick evolution of a pair
of coupled scalar quantities:
\begin{subequations}
\begin{align}
\theta_{n+1}(\beta) =& \theta_n(\beta) + 
\sgn(\epsilon)\mathcal{J}_{n+1}(\beta), 
\label{Eq:HamiltonEq1} \\
\mathcal{J}_{n+1}(\beta) =& \mathcal{J}_{n}(\beta) - \tilde{k} \sin
\theta_n(\beta) - 
 \sgn(\epsilon)2\pi\Omega,
\label{Eq:HamiltonEq2}
\end{align}
\label{Eq:Hamilton}
\end{subequations}
where $\theta(\beta)$ and $\mathcal{J}(\beta)$ can be considered a pair of
canonically conjugate classical action-angle variables.

Quantum accelerator modes are explained by stable periodic orbits in the 
$\theta(\beta)$ $\mathcal{J}(\beta)$ phase space \cite{Fishman2002}. It is
important to note that the dynamics described by Eq.\ (\ref{Eq:Hamilton}) are
{\it independent\/} of the value of $\beta$, and that this is the reason we only
need consider one phase space. We retain the $\beta$ argument due to the fact that if one
begins with an initial cloud of atoms where all $\beta$ subspaces are populated, then in
general the inital conditions propagated by Eq.\ (\ref{Eq:Hamilton}) are different, in a
$\beta$-dependent way. It is
therefore necessary to distinguish these parallel evolutions if one wishes to finally 
produce a physically meaningful answer. 

The true dependence of the dynamics on the value
of $\beta$ has been removed by Eq.\ (\ref{Eq:Transformation}), which changes the frame
in which the dynamics are observed from one falling freely with gravity, to one
which is explicitly dependent on the quasimomentum $\beta$. To determine the
dynamics of a cloud of falling atoms, it is therefore necessary tranform back
the results of Eq.\ (\ref{Eq:Hamilton}) using Eq.\ (\ref{Eq:Transformation}), so
that all quasimomentum subspaces are once again on an equal footing, i.e.,
simply falling freely with gravity.

Measurements in the Oxford experiment are thus more 
closely related to the $\mathcal{I}(\beta)$ distributions, the evolutions of 
which are
explicitly $\beta$-dependent [see Eq.\ (\ref{Eq:ProjHeisenbergMap})], than the 
$\mathcal{J}(\beta)$ distributions, as will be explained in more detail in 
Section \ref{Sec:LinkTo} \cite{Fishman2002}.

\subsubsection{Periodic orbits in the $\epsilon$-classical phase space}

The $\theta(\beta)$, $\mathcal{J}(\beta)$ phase space structure described by 
Eq.\ (\ref{Eq:Hamilton}) is $2\pi$ periodic in $\mathcal{J}(\beta)$, meaning 
that although the phase space is cylindrical [cyclic in the $\theta(\beta)$ 
angle variable and infinite in the $\mathcal{J}(\beta)$ action variable], it 
can be divided up into structurally equivalent phase-space cells, $2\pi$ long 
in the $\mathcal{J}(\beta)$ direction.

We now consider an initial condition 
$\mathcal{J}(\beta)= \tilde{\mathcal{J}}$,
$\theta(\beta)= \tilde{\theta}$, that lies exactly on a {\it periodic
orbit\/} in phase-space. By definition the momentum of such an initial condition
evolves as
\begin{equation}
\mathcal{J}_{n\mathfrak{p}}(\beta) = \tilde{\mathcal{J}} + 2\pi
n\mathfrak{j},
\label{Eq:Periodic}
\end{equation}
where $\mathfrak{p}$ is the {\it order\/} of the periodic orbit (the number of
points belonging to the periodic orbit contained within a single phase-space 
cell), and $\mathfrak{j}$ is the {\it jumping index\/} [the number of 
phase-space cells traversed after $\mathfrak{p}$ iterations of Eq.\
(\ref{Eq:Hamilton})] \cite{Note:Winding}.

\begin{figure}[tbp]
\begin{center}
\psfrag{JJ}{$\mathcal{J}$}
\includegraphics[width=3.4in,clip]{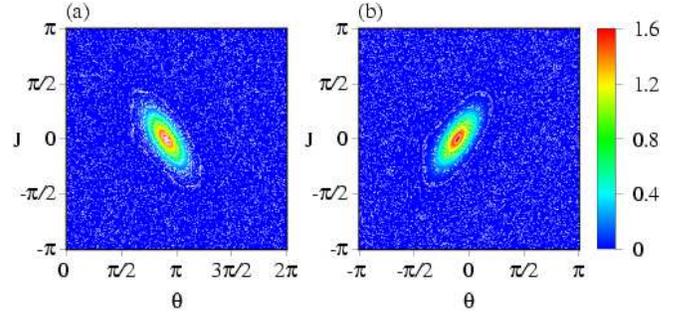}
\end{center}
\caption{(color online). Poincar\'{e} sections determined by 
 Eq.\ (\ref{Eq:Hamilton}) (white dots), 
superimposed on Wigner functions \cite{NoteWigner}
corresponding to single wavepackets of the form given in 
Eq.\ (\ref{Eq:GaussAnsatz}) (color density plots). Means and variances 
determined by Eq.\ (\ref{Eq:UC2CM}), for $\tilde{k}=2$ and (a) $\epsilon=-0.2$, 
(b) $\epsilon=0.2$. Units are dimensionless.}
\label{Fig:Poincare}
\end{figure}

If the periodic orbit is {\it stable\/} or {\it elliptic\/} in nature 
\cite{Lichtenberg1992},
then there is a finite area of phase-space surrounding the periodic orbit with
very similar dynamics, i.e., an equivalent predictable, periodic motion through
phase-space, modulated by an approximately harmonic oscillation around the
actual periodic orbit. The parameter regimes under which such 
dynamics occur can readily be determined by examining
stroboscopic Poincar\'{e} sections, produced by plotting the solutions of 
Eq.\ (\ref{Eq:Hamilton}) [modulo $2\pi$ in $\mathcal{J}(\beta)$], for a 
large number of initial conditions and values of $n$, on the same axes. If, as 
in Fig.\ \ref{Fig:Poincare}, noticable island structures, centered
on the points making up a stable periodic orbit, are apparent, 
then the action dynamics 
experienced by a significant fraction of the phase-space are approximately 
described by Eq.\ (\ref{Eq:Periodic}).

\subsubsection{Prediction of quantum accelerator modes in atom optics}
\label{Sec:LinkTo}

With Eqs.\ (\ref{Eq:Transformation})
and (\ref{Eq:Expectation}), Eq.\ (\ref{Eq:Periodic})
can be transformed into an equivalent expression for 
$\mathcal{I}_{n\mathfrak{p}}(\beta)$. 
\begin{equation}
\mathcal{I}_{n\mathfrak{p}}(\beta) = \tilde{\mathcal{I}}(\beta) + 2\pi
n[\mathfrak{j}
+\mbox{sgn}(\epsilon)\mathfrak{p}\Omega
].
\end{equation}
As each of the $\beta$ subspaces is now in an equivalent, freely falling with
gravity, $\beta$-independent frame, the value of 
$\tilde{\mathcal{I}}(\beta)$ is, unlike $\tilde{\mathcal{J}}$, 
explicitly dependent on the value of the quasimomentum $\beta$.

If the initial phase-space distribution
covers at least one phase-space cell [$2\pi\times2\pi$ in $\theta(\beta)\times
\mathcal{J}(\beta)$ space], then each of the islands surrounding a given stable
periodic orbit are populated. We can then describe an averaged evolution of
$\mathcal{I}_{n}(\beta)$. Furthermore, if the initial phase-space distribution
is centered around $\mathcal{I}(\beta)=0$ and has a finite width in
$\mathcal{I}(\beta)$, with increasing $n$, the value of
$\tilde{\mathcal{I}}(\beta)$ can be considered progressively more negligible.

Under these conditions, which are generally applicable to the atom-optics
experiments carried out at Oxford \cite{Oberthaler1999,Godun2000,dArcy2001,
Schlunk2003a,Schlunk2003b,dArcy2003,Ma2004}, the momentum evolution of those initial
conditions located within the island structures centered on a particular
$(\mathfrak{p},\mathfrak{j})$ periodic orbit is given approximately by
\begin{equation}
\mathcal{I}_{n}(\beta) \simeq 2\pi
n\left[\frac{\mathfrak{j}}{\mathfrak{p}}
+\mbox{sgn}(\epsilon)\Omega
\right].
\end{equation}
As in this approximate formulation all $\beta$-dependence of the evolution has
been removed, it is a straightforward matter to produce an equivalent expression
for the mean change in momentum of cold atoms (in a freely falling frame)
\begin{equation}
p_{n} \simeq 2\pi
n\left[\frac{\mathfrak{j}}{\mathfrak{p}}
+\mbox{sgn}(\epsilon)\Omega
\right]
\frac{\hbar G}{|\epsilon|}.
\label{Eq:ExperimentPredict}
\end{equation}

It is enhancements of the population of the observed momentum distributions of
cold cesium atoms,
near these particular values, that constitute quantum accelerator modes, 
as seen in the Oxford atom-optical experiments \cite{Oberthaler1999,Godun2000,
dArcy2001,Schlunk2003a,Schlunk2003b,dArcy2003,Ma2004}.

\section{Second-order
cumulant analysis of quantum accelerator modes}

\subsection{Derivation of the second-order cumulant map}
\label{Sec:Second-Order}

\subsubsection{Overview}

The quantities we consider are the mean angle 
$\theta(\beta) = \ave{\hat{\theta}}(\beta)$, 
the mean action 
$\mathcal{J}(\beta) = \ave{\hat{\mathcal{J}}}(\beta)$, 
the angle variance 
$\sigma^{2}(\beta) = \ave{\hat{\theta}^{2}}(\beta) - \theta(\beta)^{2}$, 
the action variance 
$S^{2}(\beta) = \ave{\hat{\mathcal{J}}^{2}}(\beta) - \mathcal{J}(\beta)^{2}$, 
and the symmetrized action-angle covariance 
$\Upsilon(\beta)= \ave{\hat{\theta}\hat{\mathcal{J}} + 
\hat{\mathcal{J}}\hat{\theta}}(\beta)/2 - \theta(\beta)\mathcal{J}(\beta)$,
corresponding to the general quantities 
$\mu_{\xi}$,
$\mu_{\zeta}$,
$\sigma_{\xi}^{2}$,
$\sigma_{\zeta}^{2}$, and
$\sigma_{\xi\zeta}^{2}$, as described in Section \ref{Sec:Gaussian},
respectively.

\subsubsection{Mean angle}

Taking the expectation value of Eq.\ (\ref{Eq:HeisenbergEq1}), 
we trivially deduce
\begin{equation}
\theta_{n+1}(\beta) = \theta_n(\beta) + \sgn(\epsilon) \mathcal{J}_{n+1}(\beta).
\label{Eq:BetaCumTheta}
\end{equation}

\subsubsection{Mean action}

Using Eq.\ (\ref{Eq:SinTheta}) to determine 
$\langle \sin\hat{\theta_{n}}\rangle (\beta)= e^{-\sigma_{n}^{2}(\beta)/2}
\sin \theta_{n}(\beta)$ (derivations for this and a number of other useful
identities are in Appendix \ref{App:Cumulants}),
it is straightforward to deduce from the expectation value of Eq.\
(\ref{Eq:HeisenbergEq2}) the corresponding equation  for the action
variable:
\begin{equation}
\mathcal{J}_{n+1}(\beta) = \mathcal{J}_n(\beta) - \tilde{k} e^{- \sigma_n^2(\beta) / 2} \sin \theta_n(\beta)
 -\sgn(\epsilon) 2\pi\Omega,
\label{Eq:BetaCumJ}
\end{equation}
which couples to the angle variance $\sigma^{2}(\beta)$. 

\subsubsection{Angle variance}

From Eqs.\
(\ref{Eq:HeisenbergEq1}) and (\ref{Eq:BetaCumTheta}),
the corresponding mapping equation for $\sigma^{2}(\beta)$ is given
by
\begin{equation}
\begin{split}
\sigma_{n+1}^{2}(\beta) =&
\ave{
\hat{\theta}_{n}^{2} +\sgn(\epsilon)(\hat{\mathcal{J}}_{n+1}\hat{\theta}_{n} 
+\hat{\mathcal{J}}_{n+1}\hat{\theta}_{n}) + \hat{\mathcal{J}}_{n+1}^{2}
}(\beta)
\\&-\theta_{n}(\beta)^{2} - 2\sgn(\epsilon)\theta_{n}(\beta)\mathcal{J}_{n+1}(\beta) -
\mathcal{J}_{n+1}(\beta)^{2}.
\end{split}
\end{equation}
Substituting in Eqs.\ (\ref{Eq:HeisenbergEq2}) and
(\ref{Eq:BetaCumJ}), we get
\begin{equation}
\begin{split}
\sigma_{n+1}^{2}(\beta) =&
\sigma_{n}^{2} (\beta)  +S_{n+1}^{2}(\beta)
\\ & + 
\sgn(\epsilon)
\langle 
\hat{\theta}_{n}\hat{\mathcal{J}}_{n} + \hat{\mathcal{J}}_{n}\hat{\theta}_{n}
-2\tilde{k}\hat{\theta}_{n}\sin\hat{\theta}_{n}
\rangle(\beta)
\\ & -2\theta_{n}(\beta)[\sgn(\epsilon)\mathcal{J}_{n}(\beta)
-\tilde{k}e^{-\sigma_{n}^{2}(\beta)}\sin\theta_{n}(\beta)].
\end{split}
\end{equation}
Finally, making use of Eq.\ (\ref{Eq:ThetaSinTheta}) to determine 
$\ave{\hat{\theta}_{n}\sin\hat{\theta}_{n}}(\beta)
-e^{-\sigma_{n}^{2}(\beta)/2}\theta_{n}(\beta)\sin\theta_{n}(\beta) =
\sigma_{n}^{2}(\beta)e^{-\sigma_{n}^{2}(\beta)/2}\cos\theta_{n}(\beta)$, 
we deduce
\begin{equation}
\begin{split}
\sigma_{n+1}^{2}(\beta) =&
\sigma_{n}^{2}(\beta) + S_{n+1}^{2}(\beta)
\\ &
+ 2\sgn(\epsilon)
[\Upsilon_{n}(\beta) - 
\tilde{k}\sigma_{n}^{2}(\beta)e^{-\sigma_{n}^{2}(\beta)/2}\cos\theta_{n}(\beta)],
\end{split}
\label{Eq:BetaCumSigma}
\end{equation}
which couples to the action variance $S^{2}(\beta)$, and to the symmetrized 
action-angle covariance
$\Upsilon(\beta)$. 

\subsubsection{Symmetrized action-angle covariance}

We first examine the covariance dynamics. From Eqs. 
(\ref{Eq:HeisenbergEq1}) and
(\ref{Eq:BetaCumTheta}),
 we deduce the mapping equation for the covariance to be
\begin{equation}
\begin{split}
\Upsilon_{n+1}(\beta) =& 
\frac{1}{2}
\langle
[\hat{\theta}_{n} + \sgn(\epsilon)\hat{\mathcal{J}}_{n+1}]
\hat{\mathcal{J}}_{n+1}\rangle(\beta)
\\ &
+\frac{1}{2}
\langle
\hat{\mathcal{J}}_{n+1}
[\hat{\theta}_{n} + \sgn(\epsilon)\hat{\mathcal{J}}_{n+1}]
\rangle (\beta)
\\ &
-[\theta_{n}(\beta) + \sgn(\epsilon)\mathcal{J}_{n+1}(\beta)]
\mathcal{J}_{n+1}(\beta).
\end{split}
\end{equation}
Substituting in Eqs.\ 
(\ref{Eq:HeisenbergEq2}) and 
(\ref{Eq:BetaCumJ}), we get
\begin{equation}
\begin{split}
\Upsilon_{n+1}(\beta) =&
\frac{1}{2}
\langle
\hat{\theta}_{n}
[\hat{\mathcal{J}}_{n}
-\tilde{k}\sin\hat{\theta}_{n}
] +
[\hat{\mathcal{J}}_{n}
-\tilde{k}\sin\hat{\theta}_{n}
]\hat{\theta}_{n}
\rangle (\beta)
\\& -\theta_{n}(\beta)[
\mathcal{J}_{n}(\beta) - \tilde{k}e^{-\sigma_{n}^{2}(\beta)/2}\sin 
\theta_{n}(\beta)
]\\ &
+  \sgn(\epsilon)S_{n+1}^{2}(\beta).
\end{split}
\end{equation}
We once again use Eq.\ (\ref{Eq:ThetaSinTheta}) to evaluate $\langle
\hat{\theta}_{n}\sin\hat{\theta}_{n} \rangle(\beta)$, finally arriving at
\begin{equation}
\begin{split}
\Upsilon_{n+1}(\beta) =& \Upsilon_{n} (\beta)
- \tilde{k}\sigma_{n}^{2}(\beta)e^{-\sigma_{n}^{2}(\beta)/2}\cos\theta_{n}(\beta)
\\&+\sgn(\epsilon)S_{n+1}^{2}(\beta).
\end{split}
\label{Eq:BetaCumUpsilon}
\end{equation}

\subsubsection{Action variance}

From Eqs. 
(\ref{Eq:HeisenbergEq2}) and
(\ref{Eq:BetaCumJ}),
 we deduce the mapping equation for the action variance to be
 \begin{equation}
\begin{split}
S_{n+1}^{2}(\beta) =&
\ave{\hat{\mathcal{J}}_{n}^{2}
+\tilde{k}^{2}\sin^{2}\hat{\theta}
-\tilde{k}(\hat{\mathcal{J}}_{n}\sin\hat{\theta}
+\sin\hat{\theta}\hat{\mathcal{J}}_{n})}(\beta)
\\ & -\mathcal{J}_{n}(\beta)^{2}
-\tilde{k}^{2}e^{-\sigma_{n}^{2}(\beta)}\sin^{2}\theta_{n}(\beta)
\\& + 2\tilde{k}\mathcal{J}_{n}(\beta)e^{-\sigma_{n}^{2}(\beta)/2}
\sin\theta_{n}(\beta).
\end{split}
\end{equation}
We make use of Eq.\ (\ref{Eq:SinSquaredTheta}) 
to determine
$
\ave{\sin^{2}\hat{\theta}(\beta)}
-e^{-\sigma_{n}^{2}}\sin^{2}\theta_{n}\beta
=
\{[1-e^{-\sigma_{n}^{2}}][1+e^{-\sigma_{n}^{2}}\cos 2 \theta (\beta)]\}/2
$,
and Eq.\ (\ref{Eq:JSinTheta}) 
to determine 
$
\ave{\hat{\mathcal{J}}_{n}(\beta)\sin\hat{\theta}(\beta)
+\sin\hat{\theta}(\beta)\hat{\mathcal{J}}_{n}(\beta)}
-\mathcal{J}_{n}(\beta)e^{-\sigma_{n}^{2}(\beta)/2}
\sin\theta_{n}(\beta)
=
e^{-\sigma_{n}^{2}(\beta)/2}
\Upsilon_{n}(\beta)\cos\theta_{n}(\beta)
$. 
Substituting these results in, we finally produce
\begin{equation}
\begin{split}
S_{n+1}^2(\beta) =& S_n^2(\beta)  - 2\tilde{k} e^{-\sigma_n^2(\beta)/2}  
\Upsilon_n(\beta) \cos \theta_n(\beta) 
\\
& + \frac{\tilde{k}^2}{2} [ 1- e^{- \sigma_n^2(\beta)}] [ 1+ e^{-
  \sigma_n^2(\beta)} \cos 2 \theta_{n}(\beta)].
\end{split}
\end{equation}

\subsubsection{Coupled mapping equations}

We thus conclude with five independent coupled mapping equations, approximately
describing the $\hat{\theta}(\beta)$, $\hat{\mathcal{J}}(\beta)$ kick-to-kick 
operator dynamics in terms of means and variances:
\begin{subequations}
\begin{align}
\theta_{n+1}(\beta) =& \theta_n(\beta) + \sgn(\epsilon) \mathcal{J}_{n+1}(\beta), 
\label{2CM1}\\
\mathcal{J}_{n+1}(\beta) =& \mathcal{J}_n(\beta) - \tilde{k} e^{- \sigma_n^2(\beta) / 2} 
\sin \theta_n(\beta) -
\sgn(\epsilon)  2\pi\Omega, \label{2CM2}\\
\begin{split}
\sigma_{n+1}^2(\beta) =& \sigma_n^2(\beta) + S_{n+1}^2(\beta) 
\\&+ 
2\sgn(\epsilon)[\Upsilon_n(\beta) 
  - \tilde{k} \sigma_n^2(\beta)  e^{- \sigma_n^2(\beta)/2} \cos
  \theta_{n}(\beta)], 
\end{split}
  \label{2CM3}\\
\begin{split}
\Upsilon_{n+1}(\beta) =& \Upsilon_n(\beta) - \tilde{k} 
\sigma_n^2 (\beta) e^{- \sigma_n^2 (\beta)/ 2} \cos 
\theta_n(\beta) 
\\&+ \sgn(\epsilon) S_{n+1}^2(\beta),
\end{split}
\label{2CM5}
\\
\begin{split}  
S_{n+1}^2(\beta) =& S_n^2(\beta)  - 2\tilde{k} e^{-\sigma_n^2(\beta)/2}  
\Upsilon_n(\beta) \cos \theta_n(\beta) 
\\
& + \frac{\tilde{k}^2}{2} [1- e^{- \sigma_n^2(\beta)}] 
[ 1+ e^{-\sigma_n^2(\beta)} \cos 2 \theta_{n}(\beta)]. 
\end{split}  \label{2CM4}
\end{align}
\label{Eq:2CM}
\end{subequations}

We note that setting all the second-order cumulants, 
$\sigma^{2}(\beta)$, $S^{2}(\beta)$, and $\Upsilon(\beta)$, to zero causes Eqs.\
(\ref{2CM1}) and (\ref{2CM2}) to revert to the effective classical mapping
described by Eqs.\ (\ref{Eq:HamiltonEq1}) and (\ref{Eq:HamiltonEq2}). In the
context of cumulant hierarchies, Eq.\ (\ref{Eq:Hamilton}) 
can be logically ordered as a first-order cumulant expansion of the dynamics of
the underlying operator-valued variables, as described by Eq.\
(\ref{Eq:Heisenberg}). The term ``$\epsilon$-classical'' is seen to be
appropriate, as it is only in the limit of the commutators (proportional to
$\epsilon$) tending to zero, that all quantum fluctuations can be summarily
neglected. The second-order cumulant map described by Eq.\ (\ref{Eq:2CM}) is
thus a lowest-order description that is still able to account for some of 
these effects.

\subsection{Periodic orbits in the second-order cumulant map}

\subsubsection{Validity of the second-order cumulant map}

When considering under what circumstances we expect the second-order cumulant
map of Eq.\ (\ref{Eq:2CM})
to be a reasonably complete description of the dynamics, we note that Gaussian
wavepackets, which are described completely by their means and variances, 
are stable when undergoing simple harmonic oscillator dynamics, and
that in the vicinity of elliptic periodic orbits, the local dynamics are
approximately harmonic \cite{Lichtenberg1992}. We thus expect Eq.\ (\ref{Eq:2CM}) to be most useful
when describing the dynamics of quantum accelerator modes, which is of course
the situation of most interest to us! 

We are thus most interested in what
occurs when the initial condition is located on top of an elliptic periodic
orbit, largely inside a stable island of the $\epsilon$-classical phase space
described by Eq.\ (\ref{Eq:Hamilton}).
If the initial condition is located in the chaotic sea of the
$\epsilon$-classical phase space, Eq.\ (\ref{Eq:2CM}) is unlikely to have much
predictive power, due to the more complicated underlying dynamics.

We note that if the angle variance remains constant
from kick to kick, i.e., $\sigma_{n+1}^{2}(\beta)=\sigma_{n}^{2}(\beta) =
\tilde{\sigma}^{2}$, then Eqs.\ (\ref{2CM1}) and (\ref{2CM2}) again
effectively revert to the $\epsilon$-classical map of Eq.\ (\ref{Eq:Hamilton}).
The only difference is that the kick strength is scaled by a Gaussian
function of the angle variance, i.e., $\tilde{k}$ is replaced by
$e^{-\tilde{\sigma}^{2}/2}\tilde{k}$. A stable solution to Eq.\ (\ref{Eq:2CM})
would therefore explain the experimentally observed presence of highly populated quantum accelerator
modes, seemingly well explained in the $\epsilon$-classical picture of Eq.\
(\ref{Eq:Hamilton}), at relatively large values of $\epsilon$.
Note also that although the $\epsilon$-classical phase-space structure 
determined by Eq.\ (\ref{Eq:Hamilton}) is clearly dependent on the value of
$\tilde{k}$ (or $e^{-\tilde{\sigma}^{2}/2}\tilde{k}$), the prediction of the
momentum evolution of those atoms occupying quantum accelerator modes
[Eq.\ (\ref{Eq:ExperimentPredict})] is not. The area of the relevant island
structure dictates the proportion of atoms to be accelerated, but not their
averaged momentum evolution. This is further confirmation of the robustness of
quantum accelerator modes as an experimentally observable effect.

\subsubsection{Example: $(\mathfrak{p},\mathfrak{j})=(1,0)$ quantum 
accelerator modes}

Notable among these are the originally discovered 
$(\mathfrak{p},\mathfrak{j})=(1,0)$ quantum accelerator modes 
around $T=T_{1/2}$
\cite{Oberthaler1999}, 
corresponding to a periodic orbit of order 1 and jumping index 0, i.e., a fixed
point, clearly observable at values of $|\epsilon|$ up to $\sim 0.8$
\cite{Schlunk2003a}.
We extend the usual procedure of determining solutions of this kind in the
$\epsilon$-classical map, where one sets 
$\theta_{n+1}(\beta) = \theta_{n}(\beta) = \tilde{\theta}$,
and
$\mathcal{J}_{n+1}(\beta) = \mathcal{J}_{n}(\beta) =
\tilde{\mathcal{J}}$, to imposing these conditions, as well as
$\sigma_{n+1}^{2}(\beta) = \sigma_{n}^{2}(\beta) = \tilde{\sigma}^{2}$,
$\Upsilon_{n+1}(\beta) = \Upsilon_{n}(\beta) = \tilde{\Upsilon}$, and
$S_{n+1}^{2}(\beta) = S_{n}^{2}(\beta) = \tilde{S}^{2}$ on Eq.\ (\ref{Eq:2CM}).
This immediately produces
\begin{subequations}
\begin{align}
\tilde{\mathcal{J}} =& 0, 
\label{FP2CM1}\\
\sin \tilde{\theta} =&  -
\sgn(\epsilon) \frac{ 2\pi\Omega}{\tilde{k}e^{- \tilde{\sigma}^2 / 2} }, 
\label{FP2CM2}\\
\begin{split}
\tilde{S}^2 =& 
- 2\sgn(\epsilon)(\Upsilon 
- \tilde{k} \tilde{\sigma}^2  e^{- \tilde{\sigma}^2/2} \cos
\tilde{\theta}), 
\end{split}
\label{FP2CM3}\\
\tilde{S}^2 =& \sgn(\epsilon) \tilde{k} 
\tilde{\sigma}^2  e^{- \tilde{\sigma}^2 / 2} \cos 
\tilde{\theta},
\label{FP2CM5}
\\
2\tilde{k} e^{-\tilde{\sigma}^2/2}  
\tilde{\Upsilon} \cos \tilde{\theta} 
 =&  \frac{\tilde{k}^2}{2} (1- e^{- \tilde{\sigma}^2}) 
( 1+ e^{-\tilde{\sigma}^2} \cos 2 \tilde{\theta}). 
\label{FP2CM4}
\end{align}
\label{Eq:FP2CM}
\end{subequations}
We note that Eqs.\ (\ref{FP2CM1}) and (\ref{FP2CM2}) can be considered identical
to the equivalent equations which would be produced when trying to determine
fixed points in the $\epsilon$-classical map, with $\tilde{k}$ replaced by 
$e^{-\tilde{\sigma}^2/2}\tilde{k}$. This follows directly from the similar result
in the second-order cumulant mapping [Eqs.\ (\ref{2CM1}) and (\ref{2CM2})].

It is now desirable to reduce these coupled equations [apart from Eq.\
the fully reduced (\ref{FP2CM1})] to closed form for the individual variables.
We begin by equating Eqs.\ (\ref{FP2CM3}) and (\ref{FP2CM5}), to determine that
\begin{equation}
\tilde{\Upsilon} =
\frac{\tilde{k}}{2}\tilde{\sigma}^{2}e^{-\tilde{\sigma}^{2}/2}
\cos\tilde{\theta}.
\end{equation}
Substituting this result into Eq.\ (\ref{FP2CM4}) we get
\begin{equation}
\tilde{\sigma}^2 e^{-\tilde{\sigma}^2}  
(1-\sin^{2} \tilde{\theta}) 
 =  \frac{1}{2} (1- e^{- \tilde{\sigma}^2}) 
[1+ e^{-\tilde{\sigma}^2}(1 -  2\sin^2 \tilde{\theta})].
\label{Eq:UpsSigmaTheta}
\end{equation}
We now substitute Eq.\ (\ref{FP2CM2}) into this, and order the result 
by powers of $e^{-\tilde{\sigma}^{2}}$. We arrive at
\begin{equation}
\begin{split}
0= & e^{-2\tilde{\sigma}^{2}}
+2\left[
\tilde{\sigma}^{2} -
\left( 
\frac{ 2\pi\Omega}{\tilde{k}}
\right)^{2}
\right]e^{-\tilde{\sigma}^{2}}
\\ &
-\left[
2(\tilde{\sigma}^{2}-1)\left( 
\frac{ 2\pi\Omega}{\tilde{k}}
\right)^{2}
+1\right],
\end{split}
\end{equation}
a closed equation for the desired fixed-point 
angle variance $\tilde{\sigma}^{2}$.
With the aid of the quadratic formula, this can be factorized to
\begin{equation}
\begin{split}
0 = &
\left\{
e^{-\tilde{\sigma}^{2}}
+
\tilde{\sigma}^{2} -
\left(
\frac{ 2\pi\Omega}{\tilde{k}}
\right)^{2}
+
\sqrt{
(\tilde{\sigma}^{2})^{2}
+
\left[
\left(
\frac{ 2\pi\Omega}{\tilde{k}}
\right)^{2}
-1
\right]^{2}
}
\right\}
\\
&\times
\left\{
e^{-\tilde{\sigma}^{2}}
+
\tilde{\sigma}^{2} -
\left(
\frac{ 2\pi\Omega}{\tilde{k}}
\right)^{2}
-
\sqrt{
(\tilde{\sigma}^{2})^{2}
+
\left[
\left(
\frac{ 2\pi\Omega}{\tilde{k}}
\right)^{2}
-1
\right]^{2}
}
\right\}.
\end{split}
\end{equation}
Thus, one of the this product of two terms must equal zero for there to be a
fixed point involving the variances as well as the mean values. That this can in
fact never occur for finite $\tilde{\sigma}^{2}$ is most easily shown 
graphically.

The quantity $( 2\pi\Omega/\tilde{k})^{2}$ can be considered a freely varying
parameter. However, from Eq.\ (\ref{FP2CM2}), we can see that for there to be a
hypothetical fixed point, then it is constrained by
\begin{equation}
0 \leq
\left(
\frac{ 2\pi\Omega}{\tilde{k}}
\right)^{2}
= e^{-\tilde{\sigma}^{2}}\sin^{2}\tilde{\theta} \leq 1,
\end{equation}
as is clearly also true in the $\epsilon$-classical limit
($e^{-\tilde{\sigma}^{2}}\rightarrow 1$).

\begin{figure}[tbp]
\begin{center}
\psfrag{ssquare}{$\tilde{\sigma}^{2}$}
\includegraphics[width=3.4in,clip]{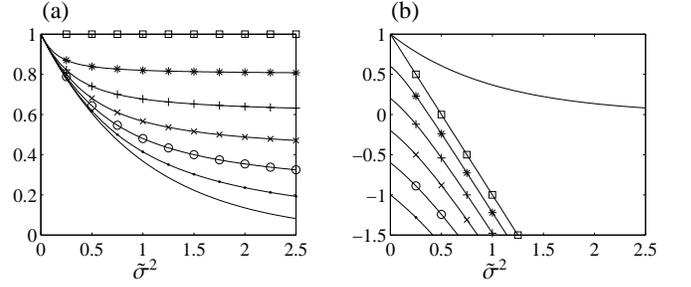}
\end{center}
\caption{Comparison of $e^{-\tilde{\sigma}^{2}}$ (solid line), with (a)
$
f_{-}(\tilde{\sigma}^{2})=
( 2\pi\Omega/\tilde{k})^{2}-\tilde{\sigma}^{2}
-
\sqrt{
(\tilde{\sigma}^{2})^{2}
+
[
( 2\pi\Omega/\tilde{k})^{2}-1
]^{2}
}
$, and (b) 
$f_{+}(\tilde{\sigma}^{2})=
( 2\pi\Omega/\tilde{k})^{2}-\tilde{\sigma}^{2}
+
\sqrt{
(\tilde{\sigma}^{2})^{2}
+
[
( 2\pi\Omega/\tilde{k})^{2}-1
]^{2}
}
$, for values of $(2\pi\Omega/\tilde{k})^{2}$ equal to 
$0$ (points), 
$0.2$ (circles),
$0.4$ (crosses),
$0.6$ (plusses),
$0.8$ (asterisks), and
$1$ (squares). Note that $f_{-}=e^{-\tilde{\sigma}^{2}}$ only when
$\tilde{\sigma}^{2}=0$, and that
$f_{+}=e^{-\tilde{\sigma}^{2}}$ only when 
$\tilde{\sigma}^{2}=0$ and $(2\pi\Omega/\tilde{k})^{2}=1$.}
\label{Fig:NoFixedPoints}
\end{figure}

The quantity $e^{-\tilde{\sigma}^{2}}$ is compared with 
$
f_{-}(\tilde{\sigma}^{2})=
( 2\pi\Omega/\tilde{k})^{2}-\tilde{\sigma}^{2}
-
\sqrt{
(\tilde{\sigma}^{2})^{2}
+
[
( 2\pi\Omega/\tilde{k})^{2}-1
]^{2}
}
$ for various values of $0\leq( 2\pi\Omega/\tilde{k})^{2}\leq 1$ 
by plotting both as functions of $\tilde{\sigma}^{2}$ in Fig.\
\ref{Fig:NoFixedPoints}(a). In
Fig.\ \ref{Fig:NoFixedPoints}(b) the same is done for 
$f_{+}(\tilde{\sigma}^{2})=
( 2\pi\Omega/\tilde{k})^{2}-\tilde{\sigma}^{2}
+
\sqrt{
(\tilde{\sigma}^{2})^{2}
+
[
( 2\pi\Omega/\tilde{k})^{2}-1
]^{2}
}
$. It can be clearly seen that both $f_{-}(\tilde{\sigma}^{2})$ and
$f_{+}(\tilde{\sigma}^{2})$ only ever intersect with $e^{-\tilde{\sigma}^{2}}$ at 
$\tilde{\sigma}^{2}=0$, which, as an extended solution cannot be considered 
reasonable in this context, is exactly the pointlike $\epsilon$-classical
limit. 

The function $e^{-\tilde{\sigma}^{2}}$ is more 
generally approximated by
$f_{-}(\tilde{\sigma}^{2})$ than $f_{+}(\tilde{\sigma}^{2})$, where
additionally $(2\pi\Omega/\tilde{k})^{2}$ must equal 1, however. This can be
seen by expanding $f_{-}(\tilde{\sigma}^{2})$ as a McLaurin series:
\begin{equation}
\begin{split}
f_{-}(\tilde{\sigma}^{2}) =&
1-\tilde{\sigma}^{2}
+\frac{(\tilde{\sigma}^{2})^{2}}{2[(2\pi\Omega/\tilde{k})^{2}-1]}
\\&
-\frac{(\tilde{\sigma}^{2})^{4}}{3[(2\pi\Omega/\tilde{k})^{2}-1]^{3}}
+\cdots,
\end{split}
\end{equation}
which is equal to $e^{-\tilde{\sigma}^{2}}$ to first order in 
$\tilde{\sigma}^{2}$, and to second order in $\tilde{\sigma}^{2}$ if 
$2\pi\Omega/\tilde{k}=0$. Thus, as shown in Fig.\ \ref{Fig:NoFixedPoints}(a),
the best matching of $f_{-}(\tilde{\sigma}^{2})$ to $e^{-\tilde{\sigma}^{2}}$
occurs near $\tilde{\sigma}^{2}=0$, and when the effect of gravity is relatively 
small.

\subsection{Constraining the second-order cumulant map to be uncertainty conserving}

\subsubsection{Motivation}

Our result on the impossibility of fixed points in the second-order cumulant map
[Eq.\ (\ref{Eq:2CM})], can be understood with the aid of some simple
considerations. Recall that finding a fixed 
point of 
$\theta(\beta)$,
$\mathcal{J}(\beta)$,
$\sigma^{2}(\beta)$,
$\Upsilon(\beta)$, and
$S^{2}(\beta)$, is 
equivalent to finding a perfectly Gaussian kick-to-kick eigenstate of the 
system. 
Even if one has an initially 
perfect Gaussian wavepacket, fully described in terms of its means and 
variances, it is not possible for it to remain Gaussian unless
no terms in the Hamiltonian are of greater than quadratic order in the
observables, or equivalently, that no terms in the corresponding Heisenberg
equations of motion are of greater than linear order.
An actual Gaussian eigenstate can only occur for the harmonic oscillator. 

Although the local
dynamics in the vicinity of an elliptic periodic orbit can be considered
approximately harmonic, the global dynamics described by both 
the  $\epsilon$-classical map [Eq.\ (\ref{Eq:Hamilton})] and the underlying
Heisenberg map  [Eq.\ (\ref{Eq:Heisenberg})]
are in fact highly nonlinear in the observables. Already at second order, the
cumulant dynamics accurately reflect this fact, and this will be true for
dynamics occuring near any $\epsilon$-classical stable periodic orbit.

\subsubsection{Enforcing uncertainty conservation}

Clearly to find solutions which are {\it approximately\/} stable fixed points in
terms of the first- and second-order cumulants, we need a way of constraining
the dynamics such that an initial perfectly Gaussian wavepacket remains Gaussian
after every iteration.

Recalling the generalized relation for Gaussian wavepackets given in Eq.\
(\ref{Eq:Uncertainty}), we can state that for a Gaussian wavepacket
$\sigma_{n+1}^{2}(\beta)S_{n+1}^{2}(\beta)-\Upsilon_{n+1}(\beta)^{2}
=\epsilon^{2}/4$. Replacing Eq.\ (\ref{2CM4}) with this constraint forcibly
slaves the dynamics of the action variance to those of the position variance and
the symmetrized action-angle covariance. In the present context, this appears to
be the simplest way to enforce genuinely Gaussian dynamics.

We substitute Eqs.\ (\ref{2CM3}) and (\ref{2CM5}) into 
$\sigma_{n+1}^{2}(\beta)S_{n+1}^{2}(\beta)-\Upsilon_{n+1}(\beta)^{2}
=\epsilon^{2}/4$, and produce,
after some straightforward manipulation:
\begin{equation}
S_{n+1}^{2}(\beta) = \frac{[\Upsilon_{n}(\beta) -
\tilde{k}\sigma_{n}^{2}(\beta)e^{-\sigma_{n}^{2}(\beta)/2}
\cos\theta_{n}(\beta)]^{2}
+\epsilon^{2}/4}
{\sigma_{n}^{2}(\beta)}.
\label{Eq:BetaGaussS}
\end{equation}
Equation (\ref{Eq:BetaGaussS}) can in turn be substituted back into Eqs.
(\ref{2CM3}) and (\ref{2CM5}), and we end up with a
closed set of four coupled mapping equations:
\begin{widetext}
\begin{subequations}
\begin{align}
\theta_{n+1}(\beta) =& \theta_n(\beta) + \sgn(\epsilon) \mathcal{J}_{n+1}(\beta), 
\label{UC2CM1}\\
{\cal{J}}_{n+1}(\beta) =& {\cal{J}}_n(\beta)
 - \tilde{k} e^{- \sigma_n^2(\beta) / 2} \sin \theta_n(\beta) -
\sgn(\epsilon)  2\pi\Omega, \label{UC2CM2}\\
\sigma_{n+1}^2(\beta) =& \sigma_n^2 (\beta) 
+ 2\sgn(\epsilon)[\Upsilon_n(\beta)  
- \tilde{k} e^{- \sigma_n^2(\beta)/2} \sigma_n^2(\beta) \cos \theta_{n}(\beta)]
+ \frac{[\Upsilon_{n}(\beta)-\tilde{k}e^{-\sigma_{n}^{2}(\beta)/2}
\sigma_{n}^{2}(\beta)\cos\theta_{n}(\beta)]^{2}
+\epsilon^{2}/4}{\sigma_{n}^{2}(\beta)},
\label{UC2CM3}
\\
\Upsilon_{n+1}(\beta) =& \Upsilon_n(\beta) - \tilde{k} e^{- \sigma_n^2(\beta) / 2} 
\sigma_n^2(\beta)  \cos 
\theta_n(\beta) 
+ \sgn(\epsilon)\frac{[\Upsilon_{n}(\beta)
-\tilde{k}e^{-\sigma_{n}^{2}(\beta)/2}\sigma_{n}^{2}(\beta)
\cos\theta_{n}(\beta)]^{2}
+\epsilon^{2}/4}{\sigma_{n}^{2}(\beta)},
\label{UC2CM4}
\end{align}
\label{Eq:UC2CM}
\end{subequations}
\end{widetext}
where $S_{n+1}^{2}(\beta)$, if desired, 
can be deduced from $\sigma_{n+1}^{2}(\beta)S_{n+1}^{2}(\beta)
-\Upsilon_{n+1}(\beta)^{2}
=\epsilon^{2}/4$. We have reduced to four independent equations in Eq.\
(\ref{Eq:UC2CM}) from the five of Eq.\ (\ref{Eq:2CM}), due to the fact that the
evolution of the action variance $S^{2}(\beta)$ is no longer independent.

\subsubsection{Fixed points}

We once again impose the conditions
$\theta_{n+1}(\beta) = \theta_{n}(\beta) = \tilde{\theta}$,
$\mathcal{J}_{n+1}(\beta) = \mathcal{J}_{n}(\beta) =
\tilde{\mathcal{J}}$, 
$\sigma_{n+1}^{2}(\beta) = \sigma_{n}^{2}(\beta) = \tilde{\sigma}^{2}$,
and 
$\Upsilon_{n+1}(\beta) = \Upsilon_{n}(\beta) = \tilde{\Upsilon}$, this time on 
Eq.\ (\ref{Eq:UC2CM}), in order to determine any fixed points. 
The condition 
$S_{n+1}^{2}(\beta) = S_{n}^{2}(\beta) = \tilde{S}^{2}$ is thus
fulfilled automatically. 

The result of imposing these conditions is given by:
\begin{subequations}
\begin{align}
 \tilde{\mathcal{J}} =& 0, 
\label{FPUC2CM1}\\
\sin \tilde{\theta} =& 
-\sgn(\epsilon)  \frac{2\pi\Omega}{e^{- \tilde{\sigma}^{2} / 2} \tilde{k}}, 
\label{FPUC2CM2}\\
\begin{split}
2\tilde{\Upsilon}  
=&   
2\tilde{k} e^{- \tilde{\sigma}^{2}/2} \tilde{\sigma}^{2} \cos \tilde{\theta}
\\&
- \sgn(\epsilon)\frac{[\tilde{\Upsilon}-\tilde{k}e^{-\tilde{\sigma}^{2}/2}
\tilde{\sigma}^{2}\cos\tilde{\theta}]^{2}
+\epsilon^{2}/4}{\tilde{\sigma}^{2}},
\end{split}
\label{FPUC2CM3}
\\
\tilde{k} e^{- \tilde{\sigma}^{2} / 2} 
\tilde{\sigma}^{2}  \cos 
\tilde{\theta}  =& 
\sgn(\epsilon)\frac{[\tilde{\Upsilon}
-\tilde{k}e^{-\tilde{\sigma}^{2}/2}\tilde{\sigma}^{2}
\cos\tilde{\theta}]^{2}
+\epsilon^{2}/4}{\tilde{\sigma}^{2}},
\label{FPUC2CM4}
\end{align}
\label{Eq:FPUC2CM}
\end{subequations}
where we note that the conditions for the mean values [Eqs.\ (\ref{FPUC2CM1})
and (\ref{FPUC2CM2})] are exactly the same as those produced in the case of the
full second-order cumulant map [Eqs.\ (\ref{FP2CM1})
and (\ref{FP2CM2})], and thus the corresponding equations in the
$\epsilon$-classical limit, with $\tilde{k}$ replaced by 
$e^{-\tilde{\sigma}^{2}/2}\tilde{k}$.

When reducing the equations for the variances to closed form, 
we expect differences to occur in the equations for the variances, however,
although
adding Eqs.\ (\ref{FPUC2CM3}) and (\ref{FPUC2CM4}) together yields
$
\tilde{\Upsilon} = 
\tilde{k} e^{- \tilde{\sigma}^{2} / 2}
\tilde{\sigma}^{2}
\cos\tilde{\theta}/2
$,
an identical result to Eq.\ (\ref{Eq:UpsSigmaTheta}).

Substituting this back into Eq.\ (\ref{FPUC2CM4}) yields
\begin{equation}
\sgn (\epsilon)\tilde{\Upsilon} =
\frac{\tilde{\Upsilon}^{2} + \epsilon^{2}/4}{2\tilde{\sigma}^{2}},
\label{Eq:UpsSigma1}
\end{equation}
which reveals that $\sgn (\epsilon)\tilde{\Upsilon}$ must be positive. 

Combining the squares of Eqs.\ (\ref{FPUC2CM2}) and (\ref{Eq:UpsSigmaTheta}),
along with the  positivity of $\sgn (\epsilon)\tilde{\Upsilon}$, yields a second
independent equation dependent only on $\tilde{\sigma}^{2}$ and 
$\tilde{\Upsilon}$:
\begin{equation}
\sgn (\epsilon)\tilde{\Upsilon} =
\sqrt{
\left(
\frac{\tilde{k}\tilde{\sigma}^{2}e^{-\tilde{\sigma}^{2}/2}}{2}
\right)^{2} -
(\pi\Omega\tilde{\sigma}^{2})^{2}
}.
\label{Eq:UpsSigma2}
\end{equation}

Substituting this back into Eq.\ (\ref{Eq:UpsSigma1}) produces
\begin{equation}
\begin{split}
2\tilde{\sigma}^{2}
\sqrt{
\left(
\frac{\tilde{k}\tilde{\sigma}^{2}e^{-\tilde{\sigma}^{2}/2}}{2}
\right)^{2} -
(\pi\Omega\tilde{\sigma}^{2})^{2}
}
= &
\left(
\frac{\tilde{k}\tilde{\sigma}^{2}e^{-\tilde{\sigma}^{2}/2}}{2}
\right)^{2} 
\\&
-
(\pi\Omega\tilde{\sigma}^{2})^{2}
+\frac{\epsilon^{2}}{4},
\end{split}
\end{equation}
a closed equation of $\tilde{\sigma}^{2}$, for which, depending on the values of
$\tilde{k}$, $ 2\pi\Omega$, and $\epsilon$, solutions do in fact exist. 
Such a Gaussian solution is thus dependent on 
$\epsilon$ as well as $\tilde{k}$ and $ 2\pi\Omega$.

This equation can therefore be used as a starting point to determine fixed 
points, for the effectively Gaussian mapping of Eq.\ (\ref{Eq:UC2CM}),
numerically. 
For the fixed points studied in this Paper,
we consider situations which 
correspond experimentally to freely varying the kicking periodicity and the 
laser intensity, with $ 2\pi\Omega=gGT^{2}$ determined by 
$T=(\epsilon+2\pi)T_{1/2}/2\pi$, $g=9.8$ ms$^{-2}$, and $G=2\pi/(447$ nm). 
For illustrative purposes, 
Wigner representations 
[
$W(\theta,\mathcal{J}) = (2\pi|\epsilon|)^{-1}\int
d\tau e^{-i\mathcal{J}\tau/|\epsilon|}
\psi^{*}(\theta-\tau/2)
\psi(\theta+\tau/2)
$
] 
\cite{Gardiner2000} 
of such ``stable'' Gaussian 
wavepackets, overlaid by Poincar\'{e} sections of the 
$\epsilon$-classical phase space \cite{Fishman2002}, are shown in Fig.\ 
\ref{Fig:Poincare}. We see that the Wigner functions 
closely match the shape of the stable island \cite{NoteWigner}. 

It should be noted that this same general procedure for finding approximate
fixed points can be applied, with slight modifications, to general periodic
orbits. A simple transformation of the action, i.e, defining 
$\hat{\mathcal{L}}_{n}(\beta)=\hat{\mathcal{J}}_{n}(\beta)-2\pi n
\mathfrak{j}/\mathfrak{p}$, causes a periodic orbit, of order $\mathfrak{p}$ 
and jumping index
$\mathfrak{j}$ in $\theta(\beta)$, $\mathcal{J}(\beta)$ space, to have jumping
index $0$ in $\theta(\beta)$, $\mathcal{L}(\beta)$ space. One can then search
for solutions of the resulting map, iterated $\mathfrak{p}$ times, in an
analogous manner.

\begin{figure}[tbp]
\begin{center}
\includegraphics[width=3.4in,clip]{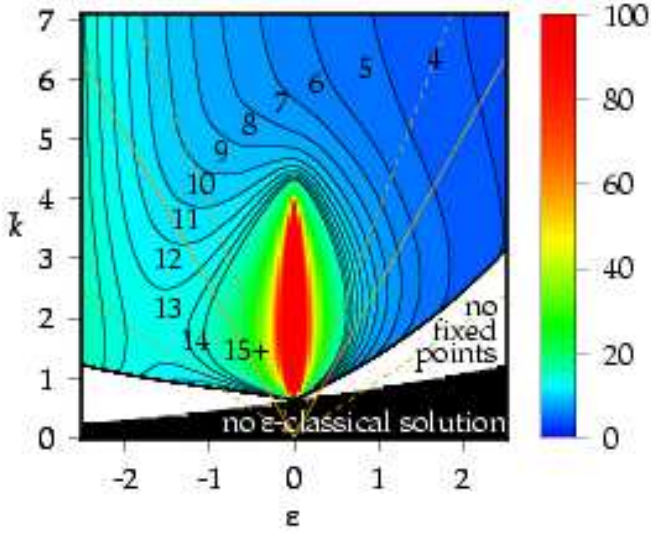}
\end{center}
\caption{(color online). Number of iterations of 
Eq.\ (\ref{Eq:2CM}) evolved by a Gaussian stable fixed point such that 
$|\mathcal{J}|<\pi$. Black indicates absence of
$\epsilon$-classical stable solutions \cite{Fishman2002}, white absence of 
Gaussian stable solutions. Numbers label the contours where $|\mathcal{J}|<\pi$ 
for that number of iterations (the number of iterations is capped at 100). The 
solid line marks the average experimental laser intensity $\phi_{d}=0.8\pi$, 
dashes demarcate its experimental range 
($0.3\pi$--$1.2\pi$) \cite{Schlunk2003a}. Units are dimensionless.}
\label{Fig:CumulantEvolution}
\end{figure}

\subsection{Exact wavepacket dynamics: Comparison}

\begin{figure}[tbp]
\begin{center}
\includegraphics[width=3.4in,clip]{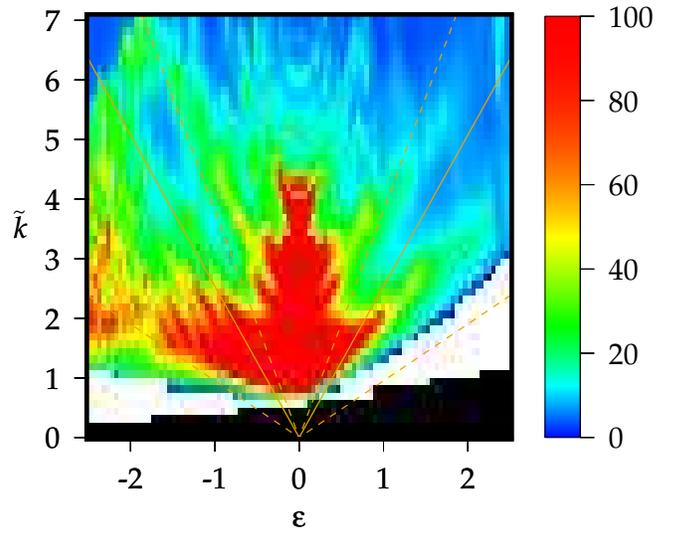}
\end{center}
\caption{(color online). As in Fig.\ \ref{Fig:CumulantEvolution}, but for exact 
wavepacket evolutions, propagating initial conditions from 
Eq.\ (\ref{Eq:ExactInitial}) with the time-evolution operator of Eq.\ 
(\ref{Eq:FloquetGauge})
($\beta=0$). Units are dimensionless.}
\label{Fig:ExactEvolution}
\end{figure}

We now come to the final point of our analysis, propagating Gaussian fixed 
point solutions using the full second-order cumulant mapping, where all 
variances must be considered explicitly.

In Fig.\ \ref{Fig:CumulantEvolution}, we 
display the time for which the center of mass momentum remains inside its 
initial phase space cell ($|\mathcal{J}_{n}(\beta)|<\pi$), using this as a 
rule-of-thumb measure of relative longevity (for a genuine fixed point this 
would be forever). We see that there is a sizable region where there there are 
no stable Gaussian fixed points, in addition to the region where there are no 
$\epsilon$-classical solutions, and that quantum accelerator modes
for $\epsilon<0$ are generally more long-lived. These observations are broadly 
born out by experiment \cite{Oberthaler1999}. 

We have also computed for how long $|\mathcal{J}_{n}|<\pi$ when integrating 
the exact evolution described by Eq.\ (\ref{Eq:FloquetGauge}) \cite{dArcy2001}.
Due to the fact that we restrict ourselves to a single $\beta$ subspace, the
wavefunction we propagate must be a Bloch state, of the form
\begin{equation}
\ave{Gz = 2\pi l + \theta |\psi} =
e^{i\beta(2\pi l + \theta)}\psi(\theta),
\end{equation}
which has a discrete momentum spectrum.
As we wish to carry out the numerics in the momentum basis,
we are therefore obliged for practical reasons to work in the 
$\hat{\mathcal{I}}(\beta)$ basis rather than that of $\hat{\mathcal{J}}$.
The reason is that the spectrum of $\hat{\mathcal{I}}=|\epsilon|\hat{k}$ is 
{\it fixed}, wheras that of $\hat{\mathcal{I}}$ {\it moves\/} with time [see Eq.
(\ref{Eq:Transformation})]. As the discreteness scale is $|\epsilon|$, this issue
vanishes in the $\epsilon$-classical limit, but it is unavoidable for any finite
value of $\epsilon$ in the fully quantum-mechanical dynamics. 

Determining the appropriate initial value for $\tilde{I}(\beta)$ from a
predetermined value of $\tilde{\mathcal{J}}$ is, with the aid of the expectation
value of Eq.\ (\ref{Eq:Transformation}), 
a straightforward matter. The values of $\tilde{\theta}$ and the variances
$\tilde{\sigma}^{2}$, $\tilde{\Upsilon}$, and $\tilde{S}^{2}$, are unaffected by
this transformation, and so we can set the initial $\psi(\theta)$ to be
\begin{equation}
\begin{split}
\psi(\theta,\beta) \propto & 
[2\pi\tilde{\sigma}^{2}]^{-1/4} 
\\&\times
\exp\left(-
\frac{[1-i2\tilde{\Upsilon}/|\epsilon|][\theta-\tilde{\theta}]^{2}}
{4\tilde{\sigma}^{2}}
+\frac{i\tilde{\mathcal{I}}(\beta)[\theta-\tilde{\theta}]}{|\epsilon|}
\right).
\label{Eq:BetaGaussInitial}
\end{split}
\end{equation}

The computationally more convenient $\mathcal{I}$ representation of the initial 
state determined by Eq.\ (\ref{Eq:FPUC2CM}) is then the discrete
Fourier transform of Eq.\ (\ref{Eq:BetaGaussInitial}), given by
$|\psi\rangle \propto \sum_{n=-\infty}^{\infty}c_{n}|\mathcal{I}=
n|\epsilon|\rangle$, where
\begin{equation}
\begin{split}
c_{n} =& \left[ 
\frac{\tilde{\sigma}^{2}}{2\pi(\epsilon^2/4+\tilde{\Upsilon}^2)}
\right]^{1/4}\\&
\times
\exp\left(
-\frac{\tilde{\sigma}^{2}/\epsilon^{2}[n|\epsilon|-\tilde{\mathcal{I}}(\beta)]^{2}}
{1-i2\tilde{\Upsilon}/|\epsilon|} 
-\frac{i\tilde{\theta}[n|\epsilon|-\tilde{\mathcal{I}}(\beta)]}{|\epsilon|}
\right).
\label{Eq:ExactInitial}
\end{split}
\end{equation}
This must in principle be normalized numerically, although for 
a Gaussian state which is well localized in $\theta$, i.e., one
where $\sqrt{\sigma^{2}}\ll 2\pi$, the effect of this will be negligible.

Figure 
\ref{Fig:ExactEvolution} shows the results of these integrations. We see 
that Fig.\ \ref{Fig:CumulantEvolution} reproduces its qualitative features 
quite well, especially for smaller values of $\epsilon$ and $\tilde{k}$. More 
surprising is the replication of a saddle-point feature at around  
$\{\epsilon=-1.5,\tilde{k}=2\}$, indicating a resurgence of stability for large 
$\epsilon$ that is clearly not an artefact of our approximations.

\section{Conclusions}

In conclusion, we have developed a general method for using second order 
cumulants to study semiclassical-like dynamics near stable periodic orbits in 
phase space. We have successfully applied this method to quantum accelerator 
mode dynamics, which operate in an unusual $\epsilon$-semiclassical regime, thus
gaining insight into the longevity of quantum accelerator modes in different 
parameter regimes. We have explictly determined the second-order
cumulant mapping for the dynamics of  the quantum $\delta$-kicked accelerator, 
taking place near integer multiples of the the half-Talbot time. In the process,
we have shown the $\epsilon$-classical dynamics derived by Fishman, Guarneri, and
Rebuzzini, to be the first order approximation within the relevant cumulant
hierarchy, we have explained why this description remains effective even for
non-negligible values of $\epsilon$, when describing quantum accelerator mode
dynamics, and we have shown why, as quantum accelerator modes are traced back to
stable periodic orbits in the relevant phase space, 
our methodology is particularly well suited for analyzing quantum acclerator
mode dynamics.

\section*{Acknowledgements}

We thank J. Cooper, R. M. Godun, I. Guarneri, T. K\"{o}hler, M. Ku\'{s}, and K. 
\.{Z}yczkowski for stimulating 
discussions. We acknowledge support from the 
ESF through BEC2000+, the UK EPSRC, 
the EU through the ``Cold Quantum Gases'' 
network, the Royal Society, the Wolfson Foundation, the Lindemann Trust,
DOE,
and NASA.

\begin{appendix}

\section{Connection between $\epsilon\rightarrow 0$ and the 
conventional semiclassical limit}
\label{App:Classical}

From Eq.\ (\ref{Eq:HamiltonPhysical}), the Heisenberg equations of motion for
$\hat{z}$ and $\hat{p}$ can be determined directly, and integrated to produce
the following operator-valued kick-to-kick map:
\begin{subequations}
\begin{align}
\hat{z}_{n+1} = & \hat{z}_{n} + \frac{T}{m} \hat{p}_{n+1} - \frac{gGT^{2}}{2}, \\
\hat{p}_{n+1} = & \hat{p}_{n} - G\hbar\phi_{d}\sin(G\hat{z}_{n}) + mgT. 
\end{align}
\label{Eq:HeisenbergConventional}
\end{subequations}
In order to reduce the number of free parameters to a minimum, we rescale the
dynamical variables to be dimensionless, such that $\hat{\chi}= G\hat{z}$ and $\hat{\rho} =
GT\hat{p}/m$. This produces, from Eq.\ (\ref{Eq:HeisenbergConventional})
\cite{dArcy2001,NoteMap},
\begin{subequations}
\begin{align}
\hat{\chi}_{n+1} = & \hat{\chi}_{n} + \hat{\rho}_{n+1} + \pi\Omega, \\
\hat{\rho}_{n+1} = & \hat{\rho}_{n} - \kappa\sin \hat{\chi}_{n} - 2\pi\Omega, 
\end{align}
\label{Eq:HeisenbergConventionalScaled}
\end{subequations}
where $\kappa = G^{2} T \hbar \phi_{d}/m$ is the
usual dimensionless stochasticity parameter associated with the (gravity-free) classical 
$\delta$-kicked rotor \cite{Lichtenberg1992}, and,
as elsewhere in this paper, $2\pi\Omega = gG^{2}T$ is a dimensionless parameter
quantifying the effect of gravity. 

The conventional semiclassical limit of these equations can be achieved by
replacing the operator-valued dynamical variables, $\hat{\chi}$ and
$\hat{\rho}$, with their expectation values, $\chi = \ave{\hat{\chi}}$ and
$\rho = \ave{\hat{\rho}}$. As explained in Section \ref{Sec:Second-Order}, this must coincide with the vanishing of the
commutator
\begin{equation}
-i[\hat{\chi},\hat{\rho}] = \kbar = \frac{\hbar G^{2} T}{m} = \frac{2\pi
T}{T_{1/2}},
\label{Eq:kbarDef}
\end{equation}
where $T_{1/2} = 2\pi m/ \hbar G^{2}$ is the half-Talbot time.

The scaled Planck constant $\kbar$ is that conventionally used in studies of
$\delta$-kicked rotor/particle dynamics, particularly when investigating the
semiclassical limit, defined as $\kbar \rightarrow 0$ \cite{Klappauf1998}. 
We recall that the
smallness parameter $\epsilon$ upon which the concept of $\epsilon$ classics and
semiclassics in this Paper is based, is defined within 
\begin{equation}
T = (2\pi\ell + \epsilon)\frac{m}{\hbar G^{2}} 
= T_{1/2}\left(\ell  + \frac{\epsilon}{2\pi}\right),
\label{Eq:EpsilonDef}
\end{equation}
where $\ell \in \mathbb{Z}$.
Substituting Eq.\ (\ref{Eq:EpsilonDef}) into Eq.\ (\ref{Eq:kbarDef}), we
therefore get
\begin{equation}
\kbar = 2\pi\ell + \epsilon,
\end{equation}
and we see that taking the $\epsilon$-classical limit is equivalent to the
conventional rescaled Planck constant $\kbar$ being equal to an integer multiple
of $2\pi$.

\section{Bloch parametrization of the dynamical variables}
\label{App:Bloch}

\subsection{Overview}
We will now explore some relevant details of our chosen parametrization of the
position and momentum variables, formulating inner products, operators, and
commutation relations in terms of the angle $\theta$, the quasimomentum $\beta$,
the discrete position $l$, and the discrete momentum $k$.

\subsection{Inner products}
We define
\begin{equation}
\ave{Gz =2\pi l + \theta| (\hbar G)^{-1}p = k + \beta}
= \frac{1}{\sqrt{2\pi}}e^{i(\beta\theta + \beta 2\pi l + k\theta)}
\label{Eq:xpInnerProduct}
\end{equation}
and, bearing in mind the limited ranges of $\theta\in [0,2\pi)$  and $\beta \in
[0,1)$, note the following identities: 
\begin{align}
\sum_{k=-\infty}^{\infty}e^{-ik(\theta-\theta')} =& 2\pi \delta(\theta-\theta'),
\label{Eq:thetadelta}\\
\int d\theta e^{i\theta(k-k')} =& 2\pi\delta_{kk'},
\label{Eq:kKdelta}\\
\sum_{l=-\infty}^{\infty}e^{-i2\pi l(\beta-\beta')} =& \delta(\beta-\beta'),
\label{Eq:betadelta}\\
\int d\beta e^{i\beta 2\pi(l-l')} =& \delta_{ll'}.
\label{Eq:lKdelta}
\end{align}

Inserting the position representation of the identity operator into the inner
product $\ave{Gz =2\pi l + \theta|Gz =2\pi l' + \theta'}$ we find, using Eq.\
(\ref{Eq:xpInnerProduct}), that
\begin{equation}
\begin{split}
\ave{Gz =2\pi l + \theta|Gz =2\pi l' + \theta'}=& 
\int d\beta 
e^{i\beta 2\pi(l-l')}
e^{i\beta(\theta-\theta')}
\\ &\times
\frac{1}{2\pi}\sum_{k=-\infty}^{\infty}
e^{ik(\theta-\theta')}.
\end{split}
\label{Eq:xxOne}
\end{equation}
Substituting Eq.\ (\ref{Eq:thetadelta}) into Eq.\ (\ref{Eq:xxOne}), 
it simplifies to
\begin{equation}
\ave{Gz =2\pi l + \theta|Gz =2\pi l' + \theta'}=
\int d\beta 
e^{i\beta 2\pi(l-l')}
\delta(\theta-\theta'),
\end{equation}
which, with the aid of Eq.\ (\ref{Eq:kKdelta}), results in the following 
general form for the inner product of two rescaled position eigenstates, in
terms of the angle $\theta$ and the discrete position variable $l$:
\begin{equation}
\ave{Gz =2\pi l + \theta|Gz =2\pi l' + \theta'}=
\delta_{ll'}\delta(\theta-\theta').
\label{Eq:xxInnerProduct}
\end{equation}

Similarly, using Eqs. (\ref{Eq:betadelta}) and (\ref{Eq:lKdelta}), one can
straightforwardly determine a corresponding expression for the inner product of
two rescaled momentum eigenstates, in terms of the quasimomentum $\beta$ and the
discrete momentum $k$:
\begin{equation}
\ave{(\hbar G)^{-1}p = k + \beta| (\hbar G)^{-1}p = k' + \beta'}
= \delta_{kk'}\delta(\beta-\beta').
\label{Eq:ppInnerProduct}
\end{equation}

\subsection{Operators}
Having determined all the relevant inner products in terms of $\theta$, $l$,
$\beta$, and $k$, we set
\begin{align}
\hat{l} = & \int d\theta \sum_{l=-\infty}^{\infty}
|Gz = 2\pi l + \theta\rangle l \langle Gz = 2\pi l + \theta|,
\label{Eq:ldef}\\
\hat{\theta} = & \int d\theta \sum_{l=-\infty}^{\infty}
|Gz = 2\pi l + \theta\rangle \theta \langle Gz = 2\pi l + \theta|,
\label{Eq:thetadef}\\
\hat{k} = & \int d\beta \sum_{k=-\infty}^{\infty}
|(\hbar G)^{-1}p = k + \beta\rangle k \langle (\hbar G)^{-1}p = k + \beta|,
\label{Eq:kdef}\\
\hat{\beta} = & \int d\beta \sum_{k=-\infty}^{\infty}
|(\hbar G)^{-1}p = k + \beta\rangle \beta \langle (\hbar G)^{-1}p = k + \beta|.
\label{Eq:betadef}
\end{align}

These operators are defined such that 
$\hat{l}|Gz = 2\pi l + \theta\rangle = l|Gz = 2\pi l + \theta\rangle$, 
$\hat{\theta}|Gz = 2\pi l + \theta\rangle = \theta|Gz = 2\pi l + \theta\rangle$, 
$\hat{k}|(\hbar G)^{-1}p = k + \beta\rangle = k|(\hbar G)^{-1}p = k +
\beta\rangle$, and
$\hat{\beta}|(\hbar G)^{-1}p = k + \beta\rangle = \beta|(\hbar G)^{-1}p = k +
\beta\rangle$, as can readily be confirmed with the aid of Eqs.\
(\ref{Eq:xxInnerProduct}) and (\ref{Eq:ppInnerProduct}). In addition, clearly
\begin{align}
\hat{\theta}+2\pi\hat{l} =& G\hat{z},
\label{Eq:PosParam}
\\
\hat{\beta}+\hat{k} =& \frac{\hat{p}}{\hbar G}.
\label{Eq:MomParam}
\end{align}

\subsection{Commutators}
We now determine the commutators of the various parametrized position and
momentum dynamical variables. 

We begin by considering the angle $\hat{\theta}$, and the quasimomentum
$\hat{\beta}$. Using Eqs.\ (\ref{Eq:thetadef}), (\ref{Eq:betadef}), and
(\ref{Eq:xpInnerProduct}), we readily determine that
\begin{equation}
\begin{split}
\hat{\beta}\hat{\theta} =& 
\frac{1}{\sqrt{2\pi}}\iint d\beta d\theta \beta \theta e^{-i\theta\beta}
\\&\times
\sum_{k,l=-\infty}^{\infty}
|(\hbar G)^{-1}p = k + \beta\rangle
e^{-i(\theta k+2\pi l \beta)}
\langle Gz = 2\pi l + \theta|.
\end{split}
\end{equation}
Sandwiching this expression between identity operators, in the position and
momentum representations, respectively, produces, 
with the aid of Eqs.\ 
(\ref{Eq:xpInnerProduct}), (\ref{Eq:thetadelta}), and (\ref{Eq:betadelta}),
\begin{equation}
\begin{split}
\hat{\beta}\hat{\theta} =& 
\frac{1}{\sqrt{2\pi}}\iint d\beta d\theta \beta \theta e^{i\theta\beta}
\\&\times
\sum_{k',l'=-\infty}^{\infty}
|(\hbar G)^{-1}p = k + \beta\rangle
e^{i\theta k'}
e^{i2\pi l' \beta}
\langle Gz = 2\pi l + \theta|
\\ =& (\hat{\beta}\hat{\theta})^{\dagger} 
\\= & \hat{\theta}\hat{\beta}.
\end{split}
\end{equation}
It therefore follows that the quasimomentum and angle operators commute, i.e.,
\begin{equation}
[\hat{\theta},\hat{\beta}] = 0.
\end{equation}

Considering now the discrete momentum $\hat{k}$ and the angle $\hat{\theta}$, we
determine, from Eqs.\  (\ref{Eq:xpInnerProduct}), (\ref{Eq:thetadef}), and
(\ref{Eq:kdef}), that
\begin{equation}
\begin{split}
\hat{k}\hat{\theta} =&
\frac{1}{\sqrt{2\pi}}\int d\theta \sum_{k=-\infty}^{\infty} k 
\theta e^{-i\theta k}
\\&\times
\int d\beta\sum_{l=-\infty}^{\infty}
|(\hbar G)^{-1}p = k + \beta\rangle
e^{-i(\theta k+2\pi l \beta)}
\langle Gz = 2\pi l + \theta|.
\end{split}
\end{equation}
In analogous fashion to what was carried out for the $\hat{\beta}\hat{\theta}$
product, we
sandwich this expression between identity operators in the position and
momentum representations, respectively, producing 
with the aid of Eq.\ (\ref{Eq:xpInnerProduct}): 
\begin{equation}
\begin{split}
\hat{k}\hat{\theta} =& 
\frac{1}{\sqrt{2\pi}}\iint d\beta' d\theta' 
\sum_{k',l'=-\infty}^{\infty}
|(\hbar G)^{-1}p = k' + \beta'\rangle
\\&\times
\frac{1}{2\pi}
e^{i[\beta'(\theta' + 2\pi l')]}
\int d\theta
\theta
e^{ik'\theta}
\sum_{k=-\infty}^{\infty}
k
e^{ik(\theta' -\theta)}
\\ &\times
\langle Gz = 2\pi l' + \theta'|.
\end{split}
\label{Eq:kthetaOne}
\end{equation}
We now use the identity \cite{Cohen1977}
\begin{equation}
\int d \theta f(\theta)\frac{\partial}{\partial \theta'}\delta(\theta-\theta')
=\frac{\partial}{\partial \theta'}f(\theta')
\end{equation}
along with Eq.\ (\ref{Eq:thetadelta})
to determine that
\begin{equation}
\begin{split}
\int d\theta
\theta
e^{ik'\theta}
\sum_{k=-\infty}^{\infty}
k
e^{ik(\theta' -\theta)}
=&
\int d\theta
\theta
e^{ik'\theta}
\left[
-i\frac{\partial}{\partial \theta'}
\sum_{k=-\infty}^{\infty}
e^{ik(\theta' -\theta)}
\right]\\
=&
-i2\pi\int d\theta
\theta
e^{ik'\theta}
\frac{\partial}{\partial \theta'}
\delta(\theta-\theta')
\\=&
2\pi(\theta'k'-i)e^{ik'\theta'},
\end{split}
\end{equation}
and insert this expression back into Eq.\ (\ref{Eq:kthetaOne}), to reveal
\begin{equation}
\hat{k}\hat{\theta} = 
(\hat{k}\hat{\theta})^{\dagger}-i = \hat{\theta}\hat{k} -i,
\end{equation}
which result is more conveniently phrased within the commutator
\begin{equation}
[\hat{\theta},\hat{k}] = i.
\end{equation}

In a very similar manner, using Eq.\ (\ref{Eq:betadelta}) and inserting a partial
derivative with respect to the quasimomentum $\beta$, it can be shown that 
$2\pi\hat{l}\hat{\beta} =
(2\pi\hat{l}\hat{\beta})^{\dagger} + i$, and therefore that
\begin{equation}
[\hat{l},\hat{\beta}] = \frac{i}{2\pi}.
\end{equation}

Referring to Eqs.\ (\ref{Eq:PosParam}) and
(\ref{Eq:MomParam}), we can now deduce from
$
[\hat{z},\hat{p}]/\hbar =
[\hat{\theta},\hat{\beta}] + [\hat{\theta},\hat{k}] + 
2\pi([\hat{l},\hat{\beta}] +
[\hat{l},\hat{k}]) = i
$ 
that
\begin{equation}
[\hat{k},\hat{l}] = \frac{i}{2\pi}.
\end{equation}

Trivially, $[\hat{\theta},\hat{l}]=[\hat{k},\hat{\beta}]=0$, and so, to
summarize:
\begin{gather}
[\hat{\theta},\hat{k}]= i, \\
[\hat{k},\hat{l}] = [\hat{l},\hat{\beta}] = \frac{i}{2\pi},\\
[\hat{\theta},\hat{\beta}] =
[\hat{\theta},\hat{l}]= 
[\hat{k},\hat{\beta}]=0.
\end{gather}

\section{Second-order cumulants}
\label{App:Cumulants}

\subsection{Counting statistics}

We choose to approximately express a given order $2n$ moment $\langle
\hat{q}_{1}\hat{q}_{2}\ldots \hat{q}_{2n}\rangle$ in terms 
of first and second order  
cumulants only, i.e., means and variances, 
setting third- and higher-order cumulants 
to zero [see Eq.\ (\ref{eq:correlation})]. The moment can be expressed as
a sum of $n$ terms, each consisting of a product of $k$ variances and
$2(n-k)$ means. 

Taking one such term, when considering how many ways there 
are to produce a product of $k$ variances and $2(n-k)$ means, this is 
equivalent
to determining how many ways there are to partition the
$\hat{q}_{i}$ terms into two subsets, one with $2k$ elements which
pairwise form variances, and the other with $2(n-k)$ elements 
individually forming means. 

Our starting point is to consider all
possible permutations (i.e., orderings) of the $\hat{q}_{i}$ terms. For
$n=3$ one possible permutation is
\begin{equation}
\hat{q}_{3}
\hat{q}_{1}
\hat{q}_{6}
\hat{q}_{2}
\hat{q}_{4}
\hat{q}_{5}.
\end{equation}
The first $2k$ terms in
the permutation are partitioned into unique neighbouring pairs, and all
subsequent terms are are partitioned into individual units, e.g.\ for $n=4$,
$k=2$ and this permutation
\begin{equation}
\langle\langle\hat{q}_{3}
\hat{q}_{1}\rangle\rangle
\langle\langle\hat{q}_{6}
\hat{q}_{2}\rangle\rangle
\langle\langle\hat{q}_{4}\rangle\rangle
\langle\langle\hat{q}_{5}\rangle\rangle.
\end{equation}
We thus have $k$
pairs, identified with variances, and $2(n-k)$ individual terms, identified with
means. There are $(2n)!$ such permutations.

As there is only one ``correct''
ordering, set by the ordering of the operators in the original moment, we must 
correct for overcounting. In
particular the ``correct'' ordering of the operators making up the variances
is in ascending order of the value of the subscript. The ordering of the
individual means and variances with respect to each other is obviously
unimportant. The complete set of equivalent orderings is thus:
\begin{equation}
\begin{split}
&\left\{\langle\langle\hat{q}_{1}
\hat{q}_{3}\rangle\rangle
\langle\langle\hat{q}_{2}
\hat{q}_{6}\rangle\rangle
\langle\langle\hat{q}_{4}\rangle\rangle
\langle\langle\hat{q}_{5}\rangle\rangle,\right.
\\ &
\langle\langle\hat{q}_{1}
\hat{q}_{3}\rangle\rangle
\langle\langle\hat{q}_{2}
\hat{q}_{6}\rangle\rangle
\langle\langle\hat{q}_{5}\rangle\rangle
\langle\langle\hat{q}_{4}\rangle\rangle,
\\ &
\langle\langle\hat{q}_{2}
\hat{q}_{6}\rangle\rangle
\langle\langle\hat{q}_{1}
\hat{q}_{3}\rangle\rangle
\langle\langle\hat{q}_{4}\rangle\rangle
\langle\langle\hat{q}_{5}\rangle\rangle,
\\ &
\langle\langle\hat{q}_{2}
\hat{q}_{6}\rangle\rangle
\langle\langle\hat{q}_{1}
\hat{q}_{3}\rangle\rangle
\langle\langle\hat{q}_{5}\rangle\rangle
\langle\langle\hat{q}_{4}\rangle\rangle,
\\ &
\langle\langle\hat{q}_{3}
\hat{q}_{1}\rangle\rangle
\langle\langle\hat{q}_{2}
\hat{q}_{6}\rangle\rangle
\langle\langle\hat{q}_{4}\rangle\rangle
\langle\langle\hat{q}_{5}\rangle\rangle,
\\ &
\langle\langle\hat{q}_{3}
\hat{q}_{1}\rangle\rangle
\langle\langle\hat{q}_{2}
\hat{q}_{6}\rangle\rangle
\langle\langle\hat{q}_{5}\rangle\rangle
\langle\langle\hat{q}_{4}\rangle\rangle,
\\ &
\langle\langle\hat{q}_{2}
\hat{q}_{6}\rangle\rangle
\langle\langle\hat{q}_{3}
\hat{q}_{1}\rangle\rangle
\langle\langle\hat{q}_{4}\rangle\rangle
\langle\langle\hat{q}_{5}\rangle\rangle,
\\ &
\langle\langle\hat{q}_{2}
\hat{q}_{6}\rangle\rangle
\langle\langle\hat{q}_{3}
\hat{q}_{1}\rangle\rangle
\langle\langle\hat{q}_{5}\rangle\rangle
\langle\langle\hat{q}_{4}\rangle\rangle,
\\ &
\langle\langle\hat{q}_{1}
\hat{q}_{3}\rangle\rangle
\langle\langle\hat{q}_{6}
\hat{q}_{2}\rangle\rangle
\langle\langle\hat{q}_{4}\rangle\rangle
\langle\langle\hat{q}_{5}\rangle\rangle,
\\ &
\langle\langle\hat{q}_{1}
\hat{q}_{3}\rangle\rangle
\langle\langle\hat{q}_{6}
\hat{q}_{2}\rangle\rangle
\langle\langle\hat{q}_{5}\rangle\rangle
\langle\langle\hat{q}_{4}\rangle\rangle,
\\ &
\langle\langle\hat{q}_{6}
\hat{q}_{2}\rangle\rangle
\langle\langle\hat{q}_{1}
\hat{q}_{3}\rangle\rangle
\langle\langle\hat{q}_{4}\rangle\rangle
\langle\langle\hat{q}_{5}\rangle\rangle,
\\ &
\langle\langle\hat{q}_{6}
\hat{q}_{2}\rangle\rangle
\langle\langle\hat{q}_{1}
\hat{q}_{3}\rangle\rangle
\langle\langle\hat{q}_{5}\rangle\rangle
\langle\langle\hat{q}_{4}\rangle\rangle,
\\ &
\langle\langle\hat{q}_{3}
\hat{q}_{1}\rangle\rangle
\langle\langle\hat{q}_{6}
\hat{q}_{2}\rangle\rangle
\langle\langle\hat{q}_{4}\rangle\rangle
\langle\langle\hat{q}_{5}\rangle\rangle,
\\ &
\langle\langle\hat{q}_{3}
\hat{q}_{1}\rangle\rangle
\langle\langle\hat{q}_{6}
\hat{q}_{2}\rangle\rangle
\langle\langle\hat{q}_{5}\rangle\rangle
\langle\langle\hat{q}_{4}\rangle\rangle,
\\ &
\langle\langle\hat{q}_{6}
\hat{q}_{6}\rangle\rangle
\langle\langle\hat{q}_{3}
\hat{q}_{1}\rangle\rangle
\langle\langle\hat{q}_{4}\rangle\rangle
\langle\langle\hat{q}_{5}\rangle\rangle,
\\ &
\left.
\langle\langle\hat{q}_{6}
\hat{q}_{2}\rangle\rangle
\langle\langle\hat{q}_{3}
\hat{q}_{1}\rangle\rangle
\langle\langle\hat{q}_{5}\rangle\rangle
\langle\langle\hat{q}_{4}\rangle\rangle
\right\}.
\end{split}
\end{equation}

In general there are 
$k!$ (all the possible orderings of the variances) multiplied by
$2^{k}$ (all the possible operator orderings within the variances to the power
of the number of variances) 
multiplied by
$[2(n-k)]!$ (all the possible orderings of the means) equivalent possibilities. 
Only one should be counted. 

\subsection{General expansions of moments in means and variances}

If the $\hat{q}_{i}$ are identical, then,  designating the mean
$\cum{\hat{q}}=\ave{\hat{q}}=\mu$ and variance
$\cum{\hat{q}^2} = \ave{\hat{q}^{2}}-\ave{\hat{q}}^{2}=\sigma^{2}$, we can
determine an approximate expression for the moment $\ave{\hat{q}^{2n}}$ purely in
terms of $\mu_{q}$ and $\sigma_{q}^{2}$: 
\begin{equation}
\ave{\hat{q}^{2n}} \simeq \sum_{k=0}^n 
\frac{(2n)!}{[2(n-k)]!k!2^{k}}
\mu^{2(n-k)} (\sigma^{2})^k.
\label{Eq:EvenMoment}
\end{equation}
For a general order $2n+1$ moment, the corresponding approximate expression
is given by:
\begin{equation}
\ave{\hat{q}^{2n+1}} \simeq \sum_{k=0}^n 
\frac{(2n+1)!}{[2(n-k)+1]!k!2^{k}}
\mu^{2(n-k)+1} (\sigma^{2})^k.
\label{Eq:OddMoment}
\end{equation}

If we consider the expectation value of an averaged symmetrized sum of products 
of a separately considered operator $\hat{q}_{1}$ with a product  of 
identical operators $\hat{q}_{2}$,
i.e., $\ave{\hat{q}_{1}\hat{q}_{2}^{2n}+\hat{q}_{2}^{2n}\hat{q}_{1}}/2$, then
Eqs.\ (\ref{Eq:EvenMoment}) and (\ref{Eq:OddMoment}) can be combined to produce
a similar approximate expression:
\begin{equation}
\begin{split}
\frac{1}{2}\ave{\hat{q}_{1}\hat{q}_{2}^{2n}+\hat{q}_{2}^{2n}\hat{q}_{1}} \simeq
& \mu_{1}\sum_{k=0}^n 
\frac{(2n)!}{[2(n-k)]!k!2^{k}}
 \mu_{2}^{2(n-k)} (\sigma_{2}^{2})^k
\\&+ 2n\sigma_{12}^{2}
\sum_{k=0}^{n-1} 
\frac{(2n-1)!}{[2(n-k)-1]!k!2^{k}}
\\&\times
\mu_{2}^{2(n-k)-1} (\sigma_{2}^{2})^k,
\end{split}
\label{Eq:EvenMixedMoment}
\end{equation}
where  $\mu_{1}=\ave{\hat{q}_{1}}$ and $\mu_{2}=\ave{\hat{q}_{2}}$ are means,
$\sigma_{2}=\ave{\hat{q}_{2}^{2}}-\ave{\hat{q}_{2}}^{2}$ is a variance, and 
$\sigma_{12}=\ave{\hat{q}_{1}\hat{q}_{2} + \hat{q}_{2}\hat{q}_{1}}/2
-\ave{\hat{q}_{1}}\ave{\hat{q}_{2}}$ is a symmetrized covariance. Similarly, 
for an equivalent symmetrized moment containing odd powers of $\hat{q}_{2}$,
\begin{equation}
\begin{split}
\frac{1}{2}\ave{\hat{q}_{1}\hat{q}_{2}^{2n+1}+\hat{q}_{2}^{2n+1}\hat{q}_{1}}  
\simeq & \mu_{1}
\sum_{k=0}^n 
\frac{(2n+1)!}{[2(n-k)+1]!k!2^{k}}
\\&\times
\mu_{2}^{2(n-k)+1} (\sigma_{2}^{2})^k
\\&+(2n+1)\sigma_{12}^{2}\sum_{k=0}^n 
\frac{(2n)!}{[2(n-k)]!k!2^{k}}
\\&\times
\mu_{2}^{2(n-k)} (\sigma_{2}^{2})^k.
\end{split}
\label{Eq:OddMixedMoment}
\end{equation}

Drawing on Eq.\ (\ref{Eq:EvenMoment}) it is now straightforward to approximately
expand the expectation value of the cosine of an operator in terms of means and 
variances:
\newpage
\begin{equation}
\begin{split}
\ave{\cos\hat{q}} 
=& \sum_{n=0}^\infty \frac{(-1)^n}{(2n)!} 
\ave{\hat{q}^{2n}} 
\\ 
\simeq&
\sum_{n=0}^\infty \frac{(-1)^n}{(2n)!}  \sum_{k=0}^n 
\frac{(2n)!}{[2(n-k)]!k!2^k} \mu^{2(n-k)} (\sigma^{2})^k 
\\ =&
 \sum_{k=0}^\infty \sum_{n=k}^\infty  \frac{(-1)^n}{[2(n-k)]! k!} 
 \mu^{2(n-k)} (\sigma^{2}/2)^k 
\\ =&
\sum_{k=0}^\infty \sum_{l=0}^\infty \frac{(-1)^{l+k}}{(2l)! k!} 
\mu^{2l} (\sigma^{2}/2)^k 
\\ =&
\sum_{k=0}^\infty \frac{(-\sigma^{2}/2)^k}{k!}   
\sum_{l=0}^\infty \frac{(-1)^l}{(2l)!}  \mu^{2l} 
\\ =&
e^{-\sigma^{2}/2} \cos \mu.
\end{split}
\label{Eq:CosMoment}
\end{equation}

One can similarly draw on Eq.\ (\ref{Eq:OddMoment}) to carry out an equivalent
expansion for the expectation value of the sine of an operator:
\begin{equation}
\begin{split}
\ave{\sin\hat{q}} 
=& 
\sum_{n=0}^\infty \frac{(-1)^n}{(2n+1)!} \ave{\hat{q}^{2n+1}} 
\\ 
\simeq&
 \sum_{k=0}^\infty \sum_{n=k}^\infty  \frac{(-1)^n}{[2(n-k)+1]! k!} 
 \mu^{2(n-k)+1} (\sigma^{2}/2)^k 
\\ =&
\sum_{k=0}^\infty \sum_{l=0}^\infty \frac{(-1)^{l+k}}{(2l+1)! k!} 
\mu^{2l+1} (\sigma^{2}/2)^k 
\\ =&
e^{- \sigma^{2}/2} \sin \mu.
\end{split}
\label{Eq:SinMoment}
\end{equation}

Finally, by making use of Eq.\ (\ref{Eq:OddMixedMoment}), one can
straightforwardly determine an approximate expansion of the expectation value of
an averaged symmetrized sum of products of the the sine of one operator
multiplied by a second operator, in terms of means and variances:
\begin{widetext}
\begin{equation}
\begin{split}
\frac{1}{2}\ave{\hat{q}_{1}\sin\hat{q}_{2} + (\sin\hat{q}_{2})\hat{q}_{1}} 
=& \sum_{n=0}^{\infty}\frac{(-1)^n}{(2n+1)!} 
\frac{1}{2}\ave{\hat{q}_{1}\hat{q}_{2}^{2n+1}+\hat{q}_{2}^{2n+1}\hat{q}_{1}}
\\ \simeq&
\sum_{n=0}^{\infty}\frac{(-1)^n}{(2n+1)!}
\sum_{k=0}^n\left\{
\mu_{1} 
\frac{(2n+1)!}{[2(n-k)+1]!k!2^{k}}
\mu_{2}^{2(n-k)+1} (\sigma_{2}^{2})^k
+\sigma_{12}^{2}
\frac{(2n+1)!}{[2(n-k)]!k!2^{k}}
\mu_{2}^{2(n-k)} (\sigma_{2}^{2})^k
\right\}
\\ =&
\sum_{k=0}^{\infty}\sum_{k=n}^{\infty}
\left\{
\mu_{1}
\frac{(-1)^{n}}{[2(n-k)+1]!k!}
\mu_{2}^{2(n-k)+1} (\sigma_{2}^{2}/2)^k
+\sigma_{12}^{2}
\frac{(-1)^{n}}{[2(n-k)]!k!}
\mu_{2}^{2(n-k)} (\sigma_{2}^{2}/2)^k\right\}
\\ =&
\sum_{k=0}^\infty \frac{(-1)^{k}}{k!} (\sigma_{2}^{2}/2)^k
\left\{\mu_{1}\sum_{l=0}^\infty \frac{(-1)^{l}}{(2l+1)!} 
\mu_{2}^{2l+1} 
+ \sigma_{12}^{2}\sum_{l=0}^\infty \frac{(-1)^{l}}{(2l)!} 
\mu_{2}^{2l}\right\}
\\ =&
e^{ - \sigma_{2}^{2}/2} (\mu_{1}  
\sin \mu_{2}+ \sigma_{12}^{2} 
\cos  \mu_{2}).
\end{split}
\label{Eq:MixedSinMoment}
\end{equation}
\end{widetext}

With these identities we now have everything necessary to determine the 
second-order cumulant dynamics quantum accelerator modes in the 
$\delta$-kicked accelerator.

\subsection{Formulation of expansions relevant to quantum accelerator mode
dynamics}

The dynamical variables relevant to $\delta$-kicked accelerator and quantum
accelerator mode dynamics are $\hat{\theta}$ and $\hat{\mathcal{J}}$.
All expectation values are evaluated within a reduced subspace particular 
to a specific value of the quasimomentum $\beta$, only, as detailed in Section
\ref{Sec:PartialTrace}.

Bearing this in mind, we determine from Eq.\ (\ref{Eq:SinMoment}) that
\begin{equation}
\ave{\sin\hat{\theta}}(\beta) = e^{-\sigma^{2}(\beta)/2}\sin\theta(\beta).
\label{Eq:SinTheta}
\end{equation}

Using  Eq.\ (\ref{Eq:CosMoment}), we determine that
\begin{equation}
\ave{\cos\hat{\theta}}(\beta) = e^{-\sigma^{2}(\beta)/2}\cos\theta(\beta),
\label{Eq:CosTheta}
\end{equation}
and that
\begin{equation}
\begin{split}
\ave{ \sin^2 \hat{\theta}}(\beta)
=& \frac{1}{2}[1 - \ave{\cos (2 \hat{\theta})}(\beta)]\\ 
=& 
\frac{1}{2}\{1 -e^{ - 2\sigma^{2}(\beta)} \cos 
[2 \theta(\beta)]\}.
\end{split}
\label{Eq:SinSquaredTheta}
\end{equation}

Using Eq.\ (\ref{Eq:MixedSinMoment}), we straightforwardly determine that
\begin{equation}
\begin{split}
\frac{1}{2}
\ave{\hat{\mathcal{J}} \sin \hat{\theta} + (\sin \hat{\theta}) \hat{\mathcal{J}}}
(\beta) =&
e^{ -\sigma^{2}(\beta)/2}\mathcal{J}(\beta)  
\sin \theta(\beta)
\\&
+ e^{ -\sigma^{2}(\beta)/2}\Upsilon(\beta) 
\cos  \theta(\beta),
\end{split}
\label{Eq:JSinTheta}
\end{equation}
and, again by using Eq.\ (\ref{Eq:MixedSinMoment}), where we set 
$\hat{q}_{1}=\hat{q}_{2}$, we determine that
\begin{equation}
\ave{\hat{\theta} \sin\hat{\theta}}(\beta) =
e^{ -\sigma^{2}(\beta)/2}[ \theta(\beta) 
\sin \theta(\beta)+ \sigma^{2}(\beta) \cos  \theta(\beta)].
\label{Eq:ThetaSinTheta}
\end{equation}

\section{Analogy: $\beta$-rotors}
\label{App:BetaRotor}

\subsection{Overview}

Subspaces of different $\beta$ (in an accelerating frame)
are decoupled in $\delta$-kicked accelerator dynamics. A wavefunction 
contained within any such subspace is periodic, multiplied by a quasimomentum
and position dependent phase $e^{-i\beta Gx}$, and can be equivalently
represented by a rotor wavefunction. We will now discuss 
the connection in some detail, in particular so as to more explicitly link the 
operator oriented
formalism used in this paper with the $\beta$-rotor picture used originally by
Fishman, Guarneri, and Rebuzzini \cite{Fishman2002}.

\subsection{Density operator}

We consider an assumed well defined and unit trace density operator, which, in the momentum representation, has the
general form
\begin{equation}
\begin{split}
\rho =& \iint d\beta d\beta' \sum_{k,k'=-\infty}^{\infty}
d_{kk'}(\beta,\beta')
\\&\times 
|(\hbar G)^{-1}p = k + \beta\rangle
\langle (\hbar G)^{-1}p = k' + \beta'|.
\label{Eq:AppGenDens}
\end{split}
\end{equation}
We select out that part of the density operator specific to a particular $\beta$
subspace by sandwiching it between $\beta$-specific projection operators
$\hat{\mathcal{P}}(\beta)$, as defined in Eq.\ (\ref{Eq:ProjectionBeta}). The
resulting expression is given by 
\begin{equation}
\begin{split}
\hat{\mathcal{P}}(\beta)\rho\hat{\mathcal{P}}(\beta)
=&\iint d\beta'' d\beta''' \sum_{k,k',k'',k'''=-\infty}^{\infty}
\delta_{k'k''}\delta(\beta-\beta'')
\\ &\times 
d_{k''k'''}(\beta'',\beta''')
\delta_{k'''k'}\delta(\beta'''-\beta)
\\ &\times 
|(\hbar G)^{-1}p = k + \beta\rangle
\langle (\hbar G)^{-1}p = k' + \beta|
\\ =&
\sum_{k,k'=-\infty}^{\infty}
d_{kk'}(\beta,\beta)
\\ &\times
|(\hbar G)^{-1}p = k + \beta\rangle
\langle (\hbar G)^{-1}p = k' + \beta|.
\label{Eq:ProjBetaRho}
\end{split}
\end{equation}
A general matrix element of this projected out 
density operator, in the position representation, is then given by the following inner product:
\begin{widetext}
\begin{equation}
\begin{split}
\sum_{k,k'=-\infty}^{\infty}
\langle Gz = 2\pi l + \theta|(\hbar G)^{-1}p = k + \beta\rangle
d_{kk'}(\beta,\beta)
\langle (\hbar G)^{-1}p = k' + \beta|
 Gz = 2\pi l + \theta\rangle 
 &=  \varrho(\theta,\theta',\beta)e^{i\beta(\theta-\theta')}e^{i2\pi(l-l')\beta},
\end{split}
\label{Eq:MatrixElementRho}
\end{equation}
\end{widetext}
where
\begin{equation}
\varrho(\theta,\theta',\beta) =
\sum_{k,k'}d_{kk'}(\beta,\beta)\frac{e^{i(k\theta-k'\theta')}}{2\pi}.
\label{Eq:ThetaDensDef}
\end{equation}
Just as for plane waves, states resulting from the projection 
$\hat{\mathcal{P}}(\beta)\rho\hat{\mathcal{P}}(\beta)$
are extended in position space, due to the
fact that they are, in general, incoherent superpositions of Bloch states for
one particular value of the quasimomentum $\beta$. 

Bloch states are not
normalizable (i.e., they lie outside the Hilbert space) when the inner product
is defined in the usual way; the trace of the expression Eq.\
(\ref{Eq:MatrixElementRho}), defined as summing and integrating over all terms
where $\theta=\theta'$ and $l=l'$, can be seen to be similarly problematical.

We note, however, that the  
normalizing factor $\mathcal{N}(\beta)$ defined in Eq.\ (\ref{Eq:PTrace})
is given by integrating over the
diagonal elements of $\varrho(\theta,\theta')$ alone, i.e.,
\begin{equation}
\begin{split}
\int d\theta \varrho(\theta,\theta,\beta) = & 
\sum_{kk'=-\infty}^{\infty}
d_{kk'}(\beta,\beta)\int d\theta \frac{e^{i(k-k')\theta}}{2\pi}
\\ = &
\sum_{k=-\infty}^{\infty}d_{kk}(\beta,\beta)
\equiv \mathcal{N}(\beta) 
\end{split}
\label{Eq:ThetaNorm}
\end{equation}
This takes advantage of the fact that integrating the diagonal matrix
elements of Eq.\ (\ref{Eq:MatrixElementRho}) over any length $2\pi$ interval
will give the same answer. This is in turn due to the periodic nature of the 
state, apart from quasimomentum dependent phases, which are in any case absent 
on the diagonal. 

Effectively, if we are restricted to a 
single $\beta$ subspace, we can
define an inner product as the integral over $\theta$ from $a$ to $a
+ 2\pi$, at a particular value of $l$, which can be chosen arbitrarily. This 
provides a well-defined norm for Bloch states, which have been mapped from the
inifinite, continuous position basis appropriate for a free particle, to the
periodic position basis appropriate for a rotor.

\subsection{Discrete momentum operator}

We consider the action of the discrete momentum operator $\hat{k}$ restricted to a
particular $\beta$ subspace, i.e., when acting on a projected density operator
as defined in Eq.\ (\ref{Eq:ProjBetaRho}). 

In the position representation, we end up with the familiar differential
operator form appropriate for dynamics occuring on a circle:
\begin{widetext}
\begin{equation}
\begin{split}
\langle Gz = 2\pi l + \theta|
\hat{k}\hat{\mathcal{P}}(\beta)\rho\hat{\mathcal{P}}(\beta)
|Gz = 2\pi l' +\theta' \rangle
=&
e^{i2\pi(l-l')\beta}
e^{i(\theta-\theta')\beta}\frac{1}{2\pi}
\sum_{k,k'=-\infty}^{\infty}
kd_{kk'}(\beta,\beta)\\
=&
e^{i2\pi(l-l')\beta}
e^{i(\theta-\theta')\beta}\
\left[
-i\frac{\partial}{\partial \theta}\varrho(\theta,\theta',\beta)
\right].
\end{split}
\end{equation}
\end{widetext}
Note that the differential operator $-i\partial/\partial \theta$ needs to be 
taken inside the quasimomentum dependent phases, and thus it acts directly on
$\varrho(\theta,\theta')$.

\subsection{$\beta$-conditional expectation values}

\subsubsection{Discrete momentum expectation value}

The (normalized) expectation value of the discrete momentum operator 
$\hat{k}$ conditioned to a single value of $\beta$
is given by taking the normalized partial trace, as defined in 
Eq.\ (\ref{Eq:PTrace}), of $\hat{k}$ multiplied by the density
operator $\rho$, i.e.,
\begin{equation}
\langle\hat{k}\rangle(\beta) = \frac{1}{\mathcal{N}(\beta)}
\sum_{k=-\infty}^{\infty}
\langle (\hbar G)^{-1}p = k + \beta|\hat{k}\rho|(\hbar G)^{-1}p = k + \beta\rangle. 
\end{equation}
Inserting the general density operator defined in Eq.\ (\ref{Eq:AppGenDens})
produces
\begin{equation}
\langle\hat{k}\rangle(\beta) = 
\frac{1}{\mathcal{N}(\beta)}
\sum_{k = -\infty}^{\infty}k d_{kk}(\beta,\beta).
\end{equation}
With the aid of Eq.\ (\ref{Eq:ThetaDensDef}),
some fairly elementary manipulations prove this expression to be fully
equivalent to
\begin{equation}
\begin{split}
\langle\hat{k}\rangle(\beta) = &
\frac{1}{\mathcal{N}(\beta)}
\int d\theta
\sum_{k,k'=-\infty}^{\infty}d_{kk'}(\beta,\beta)k
\frac{e^{i(k-k')\theta'}}{2\pi}
\\ = & \frac{1}{\mathcal{N}(\beta)}
\int d\theta
\sum_{k,k'=-\infty}^{\infty}
d_{kk'}(\beta,\beta)
\frac{1}{2\pi}
\left[
\left.
-i\frac{\partial}{\partial \theta}
e^{i(k\theta-k'\theta')}
\right|_{\theta'=\theta}
\right]
\\ =  & 
\int d\theta
\left[
\left.
-i\frac{\partial}{\partial \theta}\bar{\varrho}(\theta,\theta',\beta)
\right|_{\theta'=\theta}
\right],
\end{split}
\label{Eq:KRhoTheta}
\end{equation}
where [taking Eq.\ (\ref{Eq:ThetaNorm}) into consideration]
\begin{equation}
\bar{\varrho}(\theta,\theta',\beta) =
\frac{\varrho(\theta,\theta',\beta)}{\mathcal{N}(\beta)} =
\frac{\varrho(\theta,\theta',\beta)}{\int d\theta \varrho(\theta,\theta,\beta)}
\label{Eq:RhoAngleNorm}
\end{equation}
is the angle-dependent
component
of the $\beta$-projected density operator of Eq.\ (\ref{Eq:ProjBetaRho}), normalized to the proportion of
population present in that $\beta$-subspace.

\subsubsection{Angle expectation value}

The corresponding expectation value for $\hat{\theta}$ conditioned to a single
value of $\beta$ is, similarly,
\begin{equation}
\langle\hat{\theta}\rangle (\beta)
= \frac{1}{\mathcal{N}(\beta)}
\sum_{k=-\infty}^{\infty}
\langle
(\hbar G)^{-1}p = k + \beta|
\hat{\theta}\rho
|(\hbar G)^{-1}p = k + \beta
\rangle,
\end{equation}
by definition. Once more we substitute in  
the density operator of Eq.\ (\ref{Eq:AppGenDens}), as well as inserting 
the position representation form of the identity operator. 

This produces 

\begin{equation}
\begin{split}
\langle\hat{\theta}\rangle (\beta) =&
\frac{1}{\mathcal{N}(\beta)}
\iint d\beta' d\theta \sum_{k,k',l = -\infty}^{\infty}
\theta
d_{k'k}(\beta',\beta)
\\ &\times
\langle
(\hbar G)^{-1}p = k + \beta|Gz = 2\pi l + \theta\rangle
\\ &\times
\langle
Gz = 2\pi l + \theta
|(\hbar G)^{-1}p = k' + \beta'
\rangle 
\\ = & \frac{1}{\mathcal{N}(\beta)}
\int d\theta \theta\sum_{k,k'=-\infty}^{\infty}
d_{k'k}(\beta,\beta)\frac{e^{i(k-k')\theta}}{2\pi}
\\ = & 
\int d\theta \theta \bar{\varrho}(\theta,\theta,\beta).
\end{split}
\label{Eq:ThetaRhoTheta}
\end{equation}
We see that, following some simple
manipulations, we have arrived at an equivalent expression to Eq.\ 
(\ref{Eq:KRhoTheta}) for the angle variable, written in terms of an integral
involving  $\bar{\varrho}(\theta,\theta',\beta)$, as defined in Eq.\
(\ref{Eq:RhoAngleNorm}), only.

\subsubsection{General expectation values}

In a similar fashion to the derivations given in Eqs.\ (\ref{Eq:KRhoTheta})
and (\ref{Eq:ThetaRhoTheta}), it is now straightforward to deduce an equivalent
general expression for $\beta$-conditional expectation values: 
\begin{equation}
\langle \hat{\theta}^{n}\hat{k}^{m}\rangle (\beta)
= \int d\theta \theta^{n}
\left[
\left.
(-i)^{m}\frac{\partial^{m}}{\partial \theta^{m}}\bar{\varrho}(\theta,\theta',\beta)
\right|_{\theta'=\theta}
\right],
\end{equation}
where alternative orderings can be determined with the aid of the commutation
relations.

We thus see that for all possible expectation values of interest,
$\bar{\varrho}(\theta,\theta',\beta)$ is a density operator containing all the
information we need. The dynamics of each $\beta$-subspace can thus be mapped
onto the dynamics of separate {\em rotors}, termed $\beta$-rotors by Fishman,
Guarneri, and Rebuzzini \cite{Fishman2002}.

\end{appendix}


\begin{thebibliography}{99}
\bibitem{CommentGaussian}
This in principle assumes both $\hat{\xi}$ and $\hat{\zeta}$ to be continuously defined 
on $(-\infty,\infty)$, although in practice this restriction can frequently be 
softened.
\bibitem{Marcienkiewicz1939}
Although cumulant expansions can be extended indefinitely, note that J. Marcienkiewicz,
Math.\ Z. {\bf 44}, 612 (1939), determined that there are no well-defined
distributions that are completely described by a finite number of greater than 2
cumulants. In this sense the truncation at second order, as used here, is
somewhat special. See, also, Ref.\ \cite{Gardiner1996}.
\bibitem{NoteRecoil}
The quantity $\hbar G$, used as the basic unit of momentum in much of this
Paper, is equal to two photon recoils.
\bibitem{Note:Winding} 
A consequence of the $\mathcal{J}(\beta)$-periodicity of
the phase-space is that it can be mapped onto a torus, in which case
$\mathfrak{j}$ can be considered a winding number.
\bibitem{NoteWigner}
As the Wigner function is only for illustrative purposes, we have simplified by allowing 
both dynamical variables to be infinite and continous for its determination.
\bibitem{NoteMap}
Equation (\ref{Eq:HeisenbergConventionalScaled}) has clearly a very similar form
to Eq.\ (\ref{Eq:Heisenberg}). While this similarity is obviously not 
coincidental, one should nevertheless  bear in mind that the interpretation of
these equations is rather different.
\end{thebibliography}
\end{document}